\documentclass[12pt,preprint]{aastex}
\usepackage{natbib}
\usepackage{url}
\usepackage{lscape}
\usepackage{float}
\usepackage{amsmath}
\usepackage{natbib}
\usepackage{graphicx}
\usepackage{amsmath}
\usepackage[pdftex]{hyperref}
\begin{document}
\title{Atmospheric Constraints on the Surface UV Environment of Mars at 3.9 Ga Relevant to Prebiotic Chemistry}
\author{Sukrit Ranjan\altaffilmark{1,2}, Robin Wordsworth\altaffilmark{3,4}, Dimitar D. Sasselov\altaffilmark{1}}

\altaffiltext{1}{Harvard-Smithsonian Center for Astrophysics, Cambridge, MA 02138, USA}
\altaffiltext{2}{60 Garden Street, Mail Stop 10, Cambridge, MA 02138, USA; sranjan@cfa.harvard.edu; 617-495-5676}
\altaffiltext{3}{Harvard Paulson School of Engineering and Applied Sciences, Harvard University, Cambridge, MA 02140, USA}
\altaffiltext{4}{Department of Earth and Planetary Sciences, Harvard University, Cambridge, MA 02140, USA}

\date{\today}

\begin{abstract}
Recent findings suggest Mars may have been a clement environment for the emergence of life, and may even have compared favorably to Earth in this regard. These findings have revived interest in the hypothesis that prebiotically important molecules or even nascent life may have formed on Mars and been transferred to Earth. UV light plays a key role in prebiotic chemistry. Characterizing the early Martian surface UV environment is key to understanding how Mars compares to Earth as a venue for prebiotic chemistry. 

Here, we present two-stream multi-layer calculations of the UV surface radiance on Mars at 3.9 Ga, to constrain the surface UV environment as a function of atmospheric state. We  explore a wide range of atmospheric pressures, temperatures and compositions, corresponding to the diversity of Martian atmospheric states consistent with available constraints. We include the effects of clouds and dust. We calculate dose rates to quantify the effect of different atmospheric states on UV-sensitive prebiotic chemistry.

We find that for normative clear-sky CO$_2$-H$_2$O atmospheres, the UV environment on young Mars is comparable to young Earth. This similarity is robust to moderate cloud cover: thick clouds ($\tau_{cloud}\geq100$) are required to significantly affect the Martian UV environment, because cloud absorption is degenerate with atmospheric CO$_2$. On the other hand, absorption from SO$_2$, H$_2$S, and dust is nondegenerate with CO$_2$, meaning if these constituents build up to significant levels, surface UV fluence can be suppressed. These absorbers have spectrally variable absorption, meaning that their presence affects prebiotic pathways in different ways. In particular, high SO$_2$ environments may admit UV fluence that favors pathways conducive to abiogenesis over pathways unfavorable to it. However, better measurements of the spectral quantum yields of these pathways are required to evaluate this hypothesis definitively.

\end{abstract}


\maketitle

\maketitle

\section{Introduction}
Recent findings suggest that young Mars may have been a clement environment for the emergence of life. Analysis of Curiosity imaging of sedimentary rock strata deposited 3.2-3.6 Ga \citep{Grotzinger2015} suggest individual lakes were stable on ancient Mars for 100-10,000 years, with fluvial features laid down over 10,000-10 million years assuming formation rates corresponding to modern Earth. Similarly, Curiosity measurements of olivine and magnetite at Yellowknife crater are consistent with aqueous conditions at near-neutral pH for thousands to hundreds of thousands of years in the Noachian, with an oxidant supply that could be an energy source \citep{Bristow2015}.  In general, the geologic evidence is compelling that liquid water, a requirement for life as we know it, was present on Mars at least transiently in the Noachian \citep{Wordsworth2016}.

The young Mars may also have been a favorable environment for prebiotic chemistry (chemistry relevant to the origin of life).  Meteorite analysis has detected boron in Martian clays, important for abiogenesis since borate minerals can stabilize ribose and catalyze other prebiotic chemistry reactions (see \citealt{Stephenson2013} and sources therein). Mars may also have enjoyed greater availability of prebiotically important phosphate than Earth \citep{Adcock2013}. Climate models suggest liquid water was transient on Mars \citep{Wordsworth2013b}, which suggests the evidence of wet/dry cycles. Such cycles are useful for prebiotic chemistry: aqueous eras are beneficial for the formation of biotic monomers, while dry eras tend to concentrate feedstock molecules and aid monomer polymerization \citep{Benner2015}, relevant to the formation of nucleotides and amino acids \citep{Patel2015}. Finally, the putative dryness of Mars and the potential acidity of its early aqueous environment owing to dissolved carbonic acid from a CO$_2$-dominated atmosphere, suggest molybdate, which is suggested to catalyze formation of prebiotically important sugars such as ribose, may have been stable on Mars \citep{Benner2015, Benner2010}. Hence, there is growing interest in the possibility that prebiotically important molecules may have been produced on Mars \citep{Benner2013}, and even the hypothesis that life may have originated on Mars and been seeded to Earth \citep{Kirschvink2002, Gollihar2014, Benner2015}. 

Ultraviolet (UV) light plays a key role in prebiotic chemistry. UV photons can dissociate molecular bonds, produce ionic species, and excite molecules. These properties mean that UV light can stress prebiotic molecules \citep{Sagan1973}, but also that UV light can power synthetic prebiotic photochemistry. UV light has been invoked in prebiotic chemistry as diverse as the origin of chirality \citep{Rosenberg2008}, the synthesis of amino acid precursors \citep{Sarker2013}, and the polymerization of RNA \citep{Mulkidjanian2003}. 

The last decade has seen breakthroughs in long-standing problems in prebiotic chemistry such as the discovery of plausible mechanisms for the abiotic formation of activated pyrimidine ribonucleotides \citep{Powner2009}, the synthesis of short (2- and 3-carbon) sugars \citep{Ritson2012}, and a reaction network generating precursors for a range of prebiotically important molecules including lipids, amino acids, and ribonucleotides \citep{Patel2015}. These pathways all require UV light to function. In experiments, line sources such as low-pressure mercury lamps with monochromatic 254 nm emission are often used to simulate the incident UV radiation. However, prebiotic UV radiance in CO$_2$-dominated terrestrial-type atmospheres should instead be characterized by access to broadband fluence \citep{Ranjan2016b}. The difference can have a significant impact on prebiotic chemistry \citep{Ranjan2016a}, and there is a growing awareness in the prebiotic community of the importance of characterizing the wavelength dependence of proposed prebiotic pathways and/or using broadband sources in simulations \citep{Rapf2016}. Consequently, it is important to constrain the UV environment on the surface of Mars at epochs relevant to potential prebiotic chemistry on a spectral (wavelength-dependent) basis.

In this work, we use a two-stream multilayer radiative transfer model to constrain the surface UV environment on young Mars (3.9 Ga).  We calculate the surface radiance as a function of solar zenith angle (SZA), surface albedo ($A$), and atmospheric composition. Our model can calculate absorption and scattering due to  8 gaseous species (CO$_2$, H$_2$O, CH$_4$, SO$_2$, H$_2$S, O$_2$, and O$_3$) and 3 particulate species (H$_2$O ice, CO$_2$ ice, and Martian dust). Earlier analyses have focused on clear-sky "case studies" for the atmospheric composition; we instead explore the full range of Martian atmospheric states consistent with available geological data and climate/photochemical modelling for the Martian atmosphere. We convolve the calculated surface radiance spectra against action spectra corresponding to two different simple photochemical reactions (one useful to prebiotic chemistry, and one detrimental) that may have been important during the era of abiogenesis, and integrate the result to compute the biologically effective dose rate (BED) and estimate the impact of these parameters on prebiotic chemistry. 

In Section~\ref{sec:background}, we discuss previous work on this topic. In Section~\ref{sec:methods}, we describe our radiative transfer model and its inputs and assumptions. Section~\ref{sec:results} presents the surface radiances calculated from our model as a function of Martian atmospheric state, and Section~\ref{sec:discussion} discusses the implications for prebiotic chemistry. Section~\ref{sec:conc} summarizes our findings.

\section{Background\label{sec:background}}
Recognizing the importance of UV light to life (though mostly in the context of a stressor), previous workers have placed constraints on the primitive Martian surface UV environment. In this section, we present a review of previous work on this topic, and discuss how our work differs from them. 

\citet{Cockell2000} calculate the Martian surface flux at 3.5 Ga assuming solar input of $0.75\times$ modern, and an atmosphere composed of 1 bar CO$_2$ and 0.1 bar N$_2$. They compute their radiative transfer in a cloud-free atmosphere using a monolayer two-stream approximation with Delta-Eddington closure. They ignore water absorption, but a dust optical depth of 0.1 is assumed, as is a Lambertian surface with surface albedo $A=0.1$. \citet{Cockell2000} report the total irradiance as a function of solar time (equator at equinox) as well as the biologically-weighted\footnote{A measure of reaction rate; see Section~\ref{sec:doserates}} irradiance for DNA inactivation and photosystem damage, but not the spectral irradiance (for the early Mars case). They find the DNA inactivation-weighted irradiance to be comparable for their models of Early Mars and Early Earth, leading them to suggest that from a UV perspective the two worlds were comparably habitable.

\citet{Ronto2003} calculate the Martian surface flux at 3.5 Ga from 200-400 nm. They assume a 1-bar CO$_2$ atmosphere overlying volatilizable surface H$_2$O, and ran it through the PHOEBE photochemical model to generate atmospheric profiles for the other molecules that would be generated. They found a significant population of spectrally absorbing O$_2$, O$_3$, and NO$_2$ would be generated, including an ozone shield comparable to the modern Earth. They evaluated radiative transfer for both a pure CO$_2$ atmosphere, as well as for an atmosphere with the trace species calculated in their photochemical model. Their UV radiative transfer models assumes pure absorption and ignores scattering. In this formulation, \citet{Ronto2003} calculated Rayleigh scattering cross-sections but treated them as absorption cross-sections. This approach strongly overestimates attenuation at scattering wavelengths. It is consequently unsurprising that \citet{Ronto2003} report strong attenuation of surface UV fluence for both model atmosphere cases. in the full photochemical model case, fluence shortward of 290 nm is completely removed, due to the buildup of an ozone layer from CO$_2$ photolysis. However, \citet{Segura2007} note that this photochemical model neglects supply of reducing gases to the atmosphere due to volcanism and sinks of oxic gases due to processes involving rainout, and including either of these effects prevents the formation of an ozone layer. Geological evidence (e.g. the Tharsis plateau) indicates that young Mars had significant volcanism, which would have prevented formation of an ozone layer of the type calculated by \citet{Ronto2003}.

\citet{Cnossen2007} calculate the Martian surface flux from $\sim3.5-4$ Ga assuming a 5-bar CO$_2$, 0.8 bar N$_2$ clear-sky atmosphere. They used shortwave observations of the solar analog $\kappa^{1}$ ceti combined with a scaled solar spectrum at longer wavelengths as their top-of-atmosphere (TOA) solar input. To calculate radiative transfer, they partition the atmosphere into 40 layers. They compute absorption using the Beer-Lambert law. To account for scattering, they compute the flux scattered in each layer, assume half of it proceeds downwards and half of it proceeds upwards, and iterate this process to the surface. This approach implicitly neglects multiple scattering and assumes a surface albedo of $0$, and hence tends to overestimate atmospheric attenuation of incoming radiation. Hence, \citet{Cnossen2007} report broadband suppression of the TOA flux by multiple orders of magnitude. 

Our work builds on these previous efforts. Like \citet{Cockell2000}, \citet{Ronto2003}, and \citet{Cnossen2007}, we consider the effects of a denser CO$_2$ atmosphere; however, we consider a broader range of surface atmospheric pressures permitted by available constraints, ranging from $2\times10^{-5}-2$ bar. We build on \citet{Cockell2000}'s use of a monolayer two-stream approach to radiative transfer by using a multiple-layer two stream model, which consequently accounts for the effects of multiple scattering. Such a treatment is essential because of the unique radiative transfer regime unveiled in thick anoxic atmospheres (e.g. multibar CO$_2$ atmospheres) at UV wavelengths, characterized by the atmosphere being simultaneously optically thick and scattering-dominated. In this regime, multiple-scattering dominates and it is critical to account for its effects in order to accurately compute surface radiation environments. As a corollary, the radiative transfer treatments of \citet{Ronto2003} and \citet{Cnossen2007} are not valid in this regime\footnote{This regime is not available on the modern Earth or Mars, due to oxic absorption in the former and a thin atmosphere in the latter.}. 

In addition to varying levels of CO$_2$, our work also explores the impact of other potential atmospheric constituents on the surface UV environment. In particular, we focus on the effect of enhanced concentrations of volcanogenic gases (e.g., SO$_2$, H$_2$S), which may have been present at elevated levels on the young Mars \citep{Halevy2007, Halevy2014} and if present could have had a dramatic effect on the surface fluence \citep{Ranjan2016b}. We also explore the radiative impact of varying levels of dust and CO$_2$ and H$_2$O clouds in the Martian atmosphere, which may have been abundant \citep{Wordsworth2013b, Halevy2014}. Finally, previous workers reported the surface flux. However, as pointed out by other workers, while the flux is the relevant quantity when computing energy deposition, when computing molecular reaction rates the spherically integrated intensity, or actinic flux, is the more relevant quantity \citep{Madronich1987, Kylling1995}. For a particle lying at the planet surface, fluence below the horizon is blocked by the surface. Therefore, we report instead the integral of the intensity field at the planet surface over the hemisphere defined by elevations $>0$ (i.e. that part of the sky not blocked by the planet surface). We term this quantity the \textit{surface radiance}. For more details, see \citet{Ranjan2016b}.

\section{Methods\label{sec:methods}}
In this section, we describe the methods used to calculate the surface UV environment of early Mars. All software associated with this project is available for validation and extension at \url{https://github.com/sukritranjan/ranjanwordsworthsasselov2016}.

\subsection{Radiative Transfer Model\label{sec:methods_rtm}}
We use a multilayer two-stream approximation to compute the 1D radiative transfer of UV light through the early Martian atmosphere. Our code is based on the radiative transfer model of \citet{Ranjan2016b}. In brief, we follow the two-stream treatment of \citet{Toon1989}, and we use Gaussian (single) quadrature to connect the diffuse intensity to the diffuse flux, since \citet{Toon1989} find Gaussian quadrature closure to be more reliable than Eddington or hemispheric mean closure at short (solar) wavelengths. We include absorption and scattering due to N$_2$, CO$_2$, H$_2$O, CH$_4$, O$_2$, O$_3$, SO$_2$, and H$_2$S. For reasons of numerical stability, we assign a ceiling on the per-layer single-scattering albedo $\omega_0$ of $1-10^{-12}$. In \citet{Ranjan2016b}, we included thermal emission from the atmosphere and surface, to enable application of our code to situations where planetary UV emission might be important. Early Mars is not such a case, so here we omit these features. 

Our model requires the user to specify the partition of the atmosphere into homogenous layers, and to provide the temperature, pressure, and composition (gaseous molar concentrations) as a function of altitude. Section~\ref{sec:atmprofile} describes our calculation of these quantities. Our model also requires the user to specify the wavelength bins over which the radiative transfer is to be computed; all spectral parameters are integrated over these wavelength bins using linear interpolation in conjunction with numerical quadrature. The user also must specify the solar zenith angle (SZA) and albedo. The albedo may be specified as either a fixed value (e.g. \citealt{Rugheimer2015}), or as a wavelength- and SZA-dependent user-determined mix of the albedos corresponding to different terrestrial physical surface media (new snow, old snow, desert, tundra, ocean). 

We take the top-of-atmosphere (TOA) flux to be the solar flux at 3.9 Ga, computed at 0.1 nm resolution from the model of \citet{Claire2012} and scaled to the Martian semimajor axis of 1.524 AU. We choose 3.9 Ga for the prebiotically-relevant era 1) because of evidence for at least transient liquid water on Mars around this time \citep{Bristow2015, Grotzinger2015, Wordsworth2016}, 2) it postdates the potentially-sterilizing Late Heavy Bombardment \citep{Maher1988, Sleep1989}, and 3) it predates the bulk of the evidence for the earliest terrestrial life (see \citealt{Ranjan2016a} and sources therein). If one hypothesizes terrestrial abiogenesis was aided by transfer of prebiotically relevant compounds from Mars \citep{Benner2013, Gollihar2014, Benner2015}, then the synthesis of these molecules and their transfer must have occurred concomitantly with the origin of life on Earth. We note that, unlike the XUV, solar output varies only modestly (within a factor of 2) from 3.5-4.1 Ga in the $>180$ nm wavelength range unshielded by atmospheric CO$_2$ or H$_2$O. Therefore, our results are insensitive to the precise choice of epoch for abiogenesis.

In \citet{Ranjan2016b}, we did not include scattering and absorption due to atmospheric particulates. However, clouds have been suggested to play a major role in Martian paleoclimate \citep{Forget1997, Colaprete2003, Wordsworth2013b}. Therefore, we updated our model to allow the user to emplace CO$_2$ and H$_2$O cloud decks of user-specified optical depth (at 500 nm) in the atmosphere. The cloud decks are assumed to uniformly span the atmospheric layers into which they are emplaced. Section~\ref{sec:miecalcs} discusses the calculation of the particulate optical parameters (per-particle cross-section $\sigma$, asymmetry parameter $g$, and $\omega_0$). We use delta-scaling with Henyey-Greenstein closure \citep{Joseph1976} to correct for the effects of highly forward-peaked particulate scattering phase functions.

The fundamental output of our code is the surface radiance as a function of wavelength. The surface radiance is the integral of the intensity field at the planet surface over the unit hemisphere defined by elevations greater than zero, i.e. the intensity field integrated over all parts of the sky not blocked by the planet surface. As we argue in \citet{Ranjan2016b}, this is the relevant quantity for calculating reaction rates of molecules at planet surfaces (as compared to the actinic fluxes for molecules suspended in the atmosphere, see \citealt{Madronich1987}). In the two-stream formalism, this quantity is $$I_{surf}=F^{\downarrow}_{N}/\mu_1+F^{dir}_N/\mu_0,$$ where $F^{\downarrow}_{N}$ is the downward diffuse flux at the planet surface, $F^{dir}_N$ is the direct flux at the planet surface, and $\mu_0=\cos(SZA)$ is the cosine of the solar zenith angle. In Gaussian quadrature for the $n=1$ (two-stream) case, $\mu_1=1/\sqrt{3}$ \citep{Toon1989}; it can be interpreted as the effective zenith angle for the diffuse flux.
 
\subsection{Atmospheric Profile \label{sec:atmprofile}}
We assume a CO$_2$-dominated Martian atmosphere at 3.9 Ga. We take the atmosphere to be fully saturated with H$_2$O. Typical calculations of Noachian climate call for steady-state local surface temperatures of $T_0\lesssim273$K across the planetary surface, and more typically $T_0\sim210-250$K \citep{Forget2013, Wordsworth2013b}.Both one- and three-dimensional calculations of Noachian climate produce global mean surface temperatures of 240 K or less \citep{Forget2013, Wordsworth2013b, Ramirez2014}.  At such cold temperatures, the H$_2$O saturation pressure is very low, and H$_2$O is a trace gas in the atmosphere. Therefore, we approximate the thermodynamic properties of the Martian atmosphere by the thermodynamic properties of CO$_2$. We  take $c_p=c_{p, CO_{2}}$ and $R=R_{CO_{2}}$, where $c_p$ and $R$ are the heat capacity at constant pressure and the specific gas constant respectively. We assume the heat capacity to be constant, with $c_p=c_p(T_0)$, where $T_0$ is the surface temperature. We calculate $c_p=c_{p, CO_{2}}$ from the Shomate relation, taking the coefficients from \citet[page 115]{PPC}. We tested the effects of permitting the heat capacity to vary with temperature, and found minimal impact on our results. 

Martian paleoclimate models have been propose that invoke effects like enhanced volcanism \citep{Halevy2014} and high H$_2$ abundance \citep{Ramirez2014} to argue for global mean temperatures in excess of 273K. However, these models also require pCO$_2$ in excess of 1 bar, meaning that H$_2$O remains a trace atmospheric constituent. Regardless, our results are insensitive to the precise thermal properties of the atmosphere because of the modest variation of the absorption cross-sections of the gases in our model with temperature at UV wavelengths.

For a given surface pressure, surface temperature ($P_0$, $T_0$), we let the temperature decrease as a dry adiabat until it reaches the CO$_2$ saturation temperature, at which point it follows the CO$_2$ saturation curve. We use the empirical saturation curve of \citet{Fanale1982}, as in \citet{Wordsworth2013b}. To avoid the need for a full radiative-convective climate model, which is tangential to our objectives in this paper, we assume a stratosphere starting at 0.1 bar, following the observation of \citet{Robinson2014} that atmospheres dominated by triatomic gases tend to become optically thin and hence radiatively dominated around that pressure. We follow other workers (e.g. \citealt{Kasting1991, Hu2012, Halevy2014}) assuming the stratosphere to be isothermal; we conduct sensitivity studies demonstrating our results are not sensitive to this assumption. 

To calculate the H$_2$O saturation pressure, we use the empirical formulation of \citet{Wagner1994} via \citet{Wagner2002} for the vapor pressure of water overlying a solid reservoir (as would be the case for $T_0<273$ K). We assume the atmosphere to be fully saturated in H$_2$O until the tropopause, and we assumed the molar concentration of water in the stratosphere to be equal to its concentration at the tropopause throughout. 

Our model requires temperature, pressure, and molar concentrations as functions of altitude. To obtain a mapping between pressure and altitude, we approximate the atmosphere as a series of 1000 layers, each individually isothermal, evenly spanning $P_0-P_0\times\exp(-10)$. We then use the equation for an isothermal atmosphere in hydrostatic equilibrium, $$P(z)/P(z_0)=\exp(-(z-z_0)/H),$$  to calculate the change in altitude across each pressure layer, and sum to obtain a mapping between $z$ and $P$. Here, $P$ is pressure, $z$ is the altitude of the layer top, $z_0$ is the altitude of the layer bottom, and $H=kT/(\mu g)$ is the scale height of the layer, with $T$ being the layer temperature, $\mu$ the mean molecular weight of the atmospheric layer, and $g$ the acceleration due to Martian gravity. We also considered numerically integrating the hydrostatic equilibrium equation, $$dP/dz=-\rho g,$$ where the density $\rho=\mu P/(kT)$ for an ideal gas, directly to obtain $z(P)$. We found this approach to agree within 1\%; we consequently elected to use the simpler isothermal partition approach for our calculation. Appendix~\ref{sec:appendix_methods} presents sample atmospheric profiles derived using our methods.

\subsection{Particulate Optical Parameters\label{sec:miecalcs}}
In this section, we discuss our calculation of the optical parameters ($\sigma$, $\omega_0$, and $g$) associated with interaction of radiation with the CO$_2$ and H$_2$O ice particles that constitute clouds.

We approximate the particles as spherical, and compute their optical parameters using Mie theory at 0.1 nm resolution, following the treatment outlined in \citet{Hansen1974}. At each wavelength, we compute $\omega_0$, $g$, and the scattering efficiency $Q_{s}$. We numerically integrate these parameters over a log normal size distribution with effective radius $r_{eff}$ and effective variance $v_{eff}$, weighted by $\pi r^2 n(r)$, where $r$ is the particle radius and n(r) is the size distribution, and demand a precision of 1\% in the distribution-averaged mean values. We obtain the per-particle total extinction cross-section by $$\sigma=(\overline{Q_{s}}/\overline{\omega_0}) G,$$ where $\overline{Q_{s}}$ is the distribution-averaged mean value of $Q_{s}$, $\omega_0$ is the distribution-averaged mean value of $\omega_0$, and $$G=\int_{r_{1}}^{r_{2}}\pi r^2 n(r) dr$$ is the "geometric cross-sectional area of particles per unit volume" \citep{Hansen1974} computed at 1 ppm precision. For our numerical integral, we integrated from $r_1=r_{eff}10^{-10v_{eff}}$ to $r_2=r_{eff}10^{4v_{eff}}$; we found $n(r)<10^{-4}$ beyond these limits.

We take the index of refraction for H$_2$O ice from the compendium of \citet{Warren2008} \footnote{accessed via \url{http://www.atmos.washington.edu/ice_optical_constants}}. We take the index of refraction for CO$_2$ ice from the compendium given in \citet{PPC}\footnote{accessed via \url{http://geosci.uchicago.edu/~rtp1/PrinciplesPlanetaryClimate/Data/WorkbookDatasets/Chapter5Data/co2i4a.rfi.txt}}. This compendium was formed by G. Hansen by subjecting the absorption spectra of \cite{Hansen1997} and \citet{Hansen2005} to Kramers-Kronig analysis to obtain self-consistent spectra of real and imaginary indices of refraction (S. Warren, private communication). We take the index of refraction for Martian dust from \citet{Wolff2009} \footnote{accessed via \url{http://spacescience.arc.nasa.gov/mars-climate-modeling-group/documents/Dust_Refractive_Indicies.txt}}. The data of \citet{Wolff2009} truncate at 263 nm. Measurements of the imaginary index of refraction down to 194 nm are available from \citet{Pang1977}. We adopt the values of \citet{Pang1977} for wavelengths $<263$ nm. \citet{Zurek1978} present a compendia of the real index of refraction of Martian dust. Their study indicates that the index of refraction changes by only $\sim0.1$ from $\sim200-263$ nm. Their values at 263 nm are 0.4 higher than \citet{Wolff2009}; to avoid a discontinuity, we subtract 0.4 from the real indices of refraction of \citet{Zurek1978}.

\begin{figure}[H]
\centering
\includegraphics[width=10 cm, angle=0]{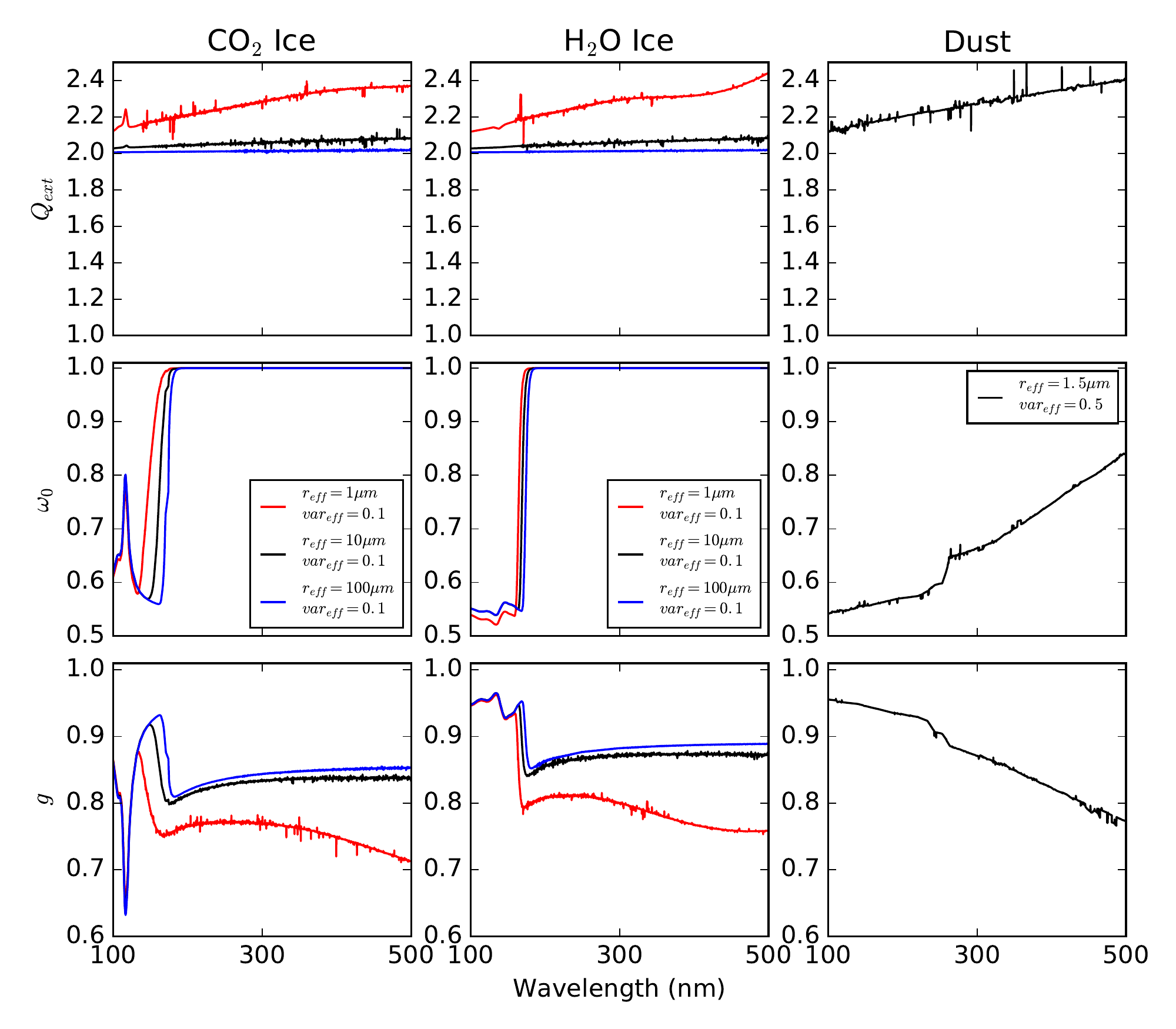}
\caption{Scattering efficiency $Q_{ext}$, single-scattering albedo $\omega_0$, and asymmetry parameter $g$ as a function of wavelength for CO$_2$ ice, H$_2$O ice, and modern Martian dust, integrated over the specified log-normal size distributions. \label{fig:mie}}
\end{figure}

Figure~\ref{fig:mie} presents the CO$_2$ and H$_2$O ice optical parameters as a function of wavelength for size distributions with $v_{eff}=0.1$ and $r_{eff}=1, 10, 100$ microns. Previous work has assumed CO$_2$ cloud particle sizes to be in the 1-100 $\mu$m range (e.g., \citealt{Forget1997}). Microphysical modelling by \citet{Colaprete2003} suggests that primitive Martian CO$_2$ clouds may have been characterized by large particle sizes, as high as $r_eff=100 \mu$m. The optical properties of CO$_2$ and H$_2$O ice particles for $r_{eff}\geq10 \mu$m insensitive to $r_{eff}$; we attribute this to size parameter $x=2\pi r_{eff}/\lambda$ being large in this regime, meaning such particles approach the large-particle limit. ($x>12$ for $r_{eff}\geq 10 \mu$m and $\lambda\leq500$ nm). For wavelengths satisfying $195<\lambda<500$ nm, CO$_2$ and H$_2$O ice are characterized by $\omega_0\approx1$. This means that CO$_2$ and H$_2$O clouds do not significantly absorb at wavelengths unshielded by H$_2$O ($<198$ nm) or CO$_2$ ($<204$ nm). By contrast, dust absorbs across $100-500$ nm, meaning dust particles can supply absorption at wavelengths not shielded by CO$_2$ or H$_2$O.

\subsection{Action Spectra and Calculation of Dose Rates\label{sec:doserates}}
To quantify the impact of different surface UV radiation environments on prebiotic chemistry, we follow the approach of \citet{Cockell1999}, \citet{Ronto2003}, and \citet{Rugheimer2015} in computing Biologically Effective Dose rates (BEDs).  Specifically, we compute $$D=(\int_{\lambda_{0}}^{\lambda_{1}} d\lambda \mathcal A(\lambda)I_{surf}(\lambda)),$$ where $\mathcal A(\lambda)$ is an action spectrum, $\lambda_0$ and $\lambda_1$ are the limits over which $\mathcal A(\lambda)$ is defined, and $I_{surf}(\lambda)$ is the UV surface radiance. An action spectrum parametrizes the relative impact of radiation on a given photoprocess as a function of wavelength, with a higher value of $\mathcal A$ meaning that a higher fraction of the incident photons are being used in the photoprocess. Hence, $D$  is proportional to the reaction rate rate of a given photoprocess for a single molecule at the surface of a planet. 

As $D$ is a relative measure of reaction rate, a normalization is required to assign a physical interpretation to its value. In this paper, we report $$\overline{D}=D/D_{\earth},$$ where $D_{\earth}$ is the dose rate on 3.9 Ga Earth. The atmospheric model for 3.9 Ga Earth is taken from \citet{Rugheimer2015}, who use a 1D coupled climate-photochemistry model to compute the atmospheric profile (T, P, composition) for the Earth at 3.9 Ga, assuming modern abiotic outgassing rates and a background atmosphere of 0.9 bar N$_2$, 0.1 bar CO$_2$, with SZA=$60^\circ$ and $A=0.2$. Consequently, $\overline{D}>1$ means that the photoprocess is proceeding faster on the Martian surface under the specified atmosphere than it would on the surface of the \citet{Rugheimer2015} fiducial Earth. Note this normalization is different from what we chose in \citet{Ranjan2016b}, because here we are trying to assess how Mars compares to the Earth as a venue for prebiotic chemistry.

Previous workers used action spectra of UV stress on modern biology (e.g. the DNA inactivation action spectrum) \citep{Cockell2000, Cockell2002, Cnossen2007, Rugheimer2015} as a gauge of the level of stress imposed by UV fluence on the prebiotic environment. However, these action spectra are based on modern life. Modern organisms have evolved sophisticated methods to deal with environmental stress, including UV exposure, that would not have been available to the first life. Further, this approach presupposes that UV light is solely a stressor, and ignores its potential role as a eustressor for abiogenesis. In this work, we follow the reasoning of our previous efforts in \citet{Ranjan2016b} in formulating action spectra corresponding to simple photoreactions that are expected to have played major roles in prebiotic chemistry. We consider two reactions: a stressor process, to capture the stress UV light places on nascent biology, and an eustressor process, to capture the role of UV light in promoting prebiotic chemistry. A detailed description of these processes and their corresponding action spectra is given in \citet{Ranjan2016b}; a brief outline is presented below.

\subsubsection{Stressor Process: Cleavage of N-Glycosidic Bond of UMP}
For our stressor process, we chose the cleavage of the N-glycosidic bond in the RNA monomer uridine monophospate (UMP). UV radiation can cleave the N-glyocosidic bond which joins the sugar to the nucleobase \citep{Gurzadyan1994}, irreversibly destroying this molecule. Hence, this process represents a stressor to abiogenesis.

The action spectrum is equal to the product of the absorption spectrum (fraction of incident photons absorbed) and the quantum yield curve (QY, fraction of absorbed photons that lead to the photoreaction). We take our UMP absorption spectrum from the work of \citep{Voet1963} (pH=7.6). Detailed spectral measurements of the QY of glyocosidic bond cleavage have not been obtained. However, \citet{Gurzadyan1994} found the QY of N-glycosidic bond cleavage in UMP in neutral aqueous solution saturated with Ar (i.e. anoxic) to be $4.3\times10^{-3}$ at 193 nm and $(2-3)\times10^{-5}$ for 254 nm. We therefore represent the QY curve as a step function with value $4.3\times10^{-3}$ for $\lambda\leq\lambda_0$ and $2.5\times10^{-5}$ for $\lambda>\lambda_0$. We consider $\lambda_0$ values of 193 and 254 nm, corresponding to the empirical limits from \citet{Gurzadyan1994}. We also consider $\lambda_0=230$ nm, which corresponds to the end of the broad absorption feature centered near 260 nm corresponding to the $\pi-\pi^{*}$ transition and also to the transition to irreversible decomposition suggested by \citet{Sinsheimer1949}. As shorthand, we refer to this photoprocess under the assumption that $\lambda_0=$X nm by UMP-X. Figure~\ref{fig:actspec} presents these action spectra.

The absorption spectra of the other RNA monomers are structurally similar to UMP \citep{Voet1963}, and the quantum yield of N-glycosidic bond cleavage in adenosine monophospate (AMP) increases at short wavelengths like UMP's does \citep{Gurzadyan1994}, leading us to argue that action spectra for N-glycosidic bond cleavage of the other RNA monomers should be broadly similar to that for UMP. Therefore, results derived using the action spectrum for UMP N-glycosidic bond cleavage should be broadly applicable to the other RNA monomers: if a UV environment is destructive for UMP, it should be bad for the other RNA monomers, and hence for abiogenesis in the RNA world hypothesis, as well.

\subsubsection{Eustressor Process: Production of Aquated Electrons from Photoionization of Cyanocuprate}
For our eustressor process, we choose the production of aquated electrons from the irradiation of a tricyanocuprate (CuCN$_3^{2-}$) complex. We chose this process because it underlies the selective 2- and 3-carbon sugar (glycolaldehyde and glyceraldehyde) synthesis pathway of \citet{Ritson2012}, which is the best candidate proposed so far for a selective prebiotic synthesis of these sugars. These sugars are required for the synthesis of RNA, and hence abiogenesis in the RNA world hypothesis. This process is also important to the prebiotic reaction network of \citet{Patel2015}. More generally, aquated electrons are useful for a broad range of reductive prebiotic chemistry, e.g., the reduction of nitriles to amines, aldehydes to hydroxyls, and hydroxyls to alkyls\footnote{J. Szostak, private communication, 2/5/16}. Therefore, this process represents a eustressor to abiogenesis. While other UV-sensitive processes conducive to abiogenesis doubtless exist, we argue this process is of particular interest because of its unique role in the most promising plausibly prebiotic pathways to the RNA monomers.

\begin{figure}[H]
\centering
\includegraphics[width=10 cm, angle=0]{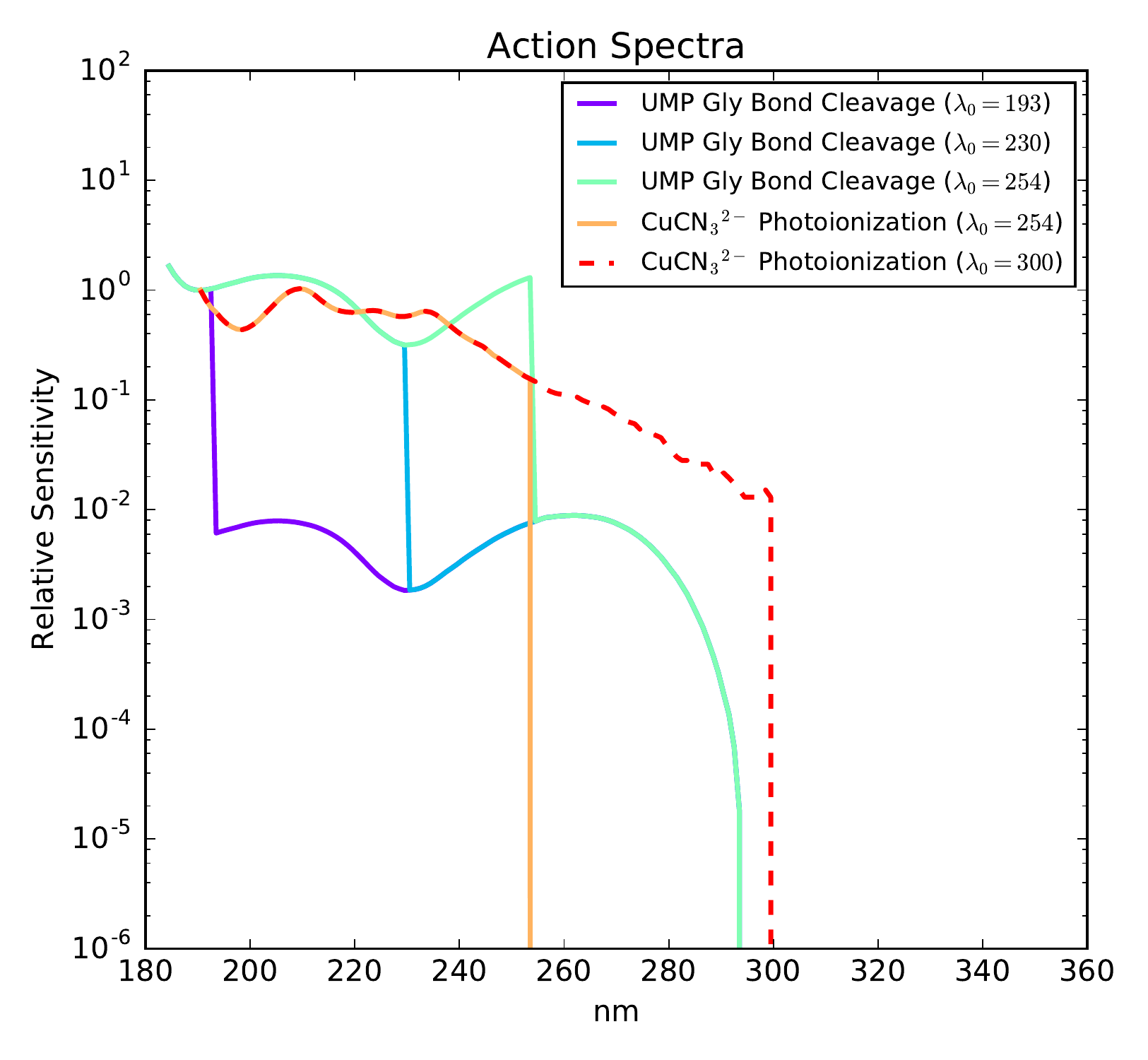}
\caption{Action spectra for photolysis of UMP$-\lambda_0$ and photoionization of CuCN3 $-\lambda_0$, assuming a step-function form to the QE for both processes with step at $\lambda_0$. The spectra are arbitrarily normalized to 1 at 190 nm. \label{fig:actspec}}
\end{figure}

We again form the action spectrum by multiplying the absorption spectrum and the quantum yield curve. We take the cyanocuprate absorption spectrum from the work of \citet{Magnani2015}, via \citet{Ranjan2016a}. The spectral QY of aquated electron production from cyanocuprate irradiation is not known. However, \citet{Horvath1984} measure a QY of 0.06 for this process at 254 nm. Following \citet{Ritson2012}'s hypothesis that photoionization of the complex drives aquated electron production, we assume the QY to be characterized by a step function with value $0.06$ for $\lambda\leq\lambda_0$ and $0$ otherwise. We empirically know $\lambda_0\geq254$ nm. To explore a range of $\lambda_0$, we consider $\lambda_0=254$ nm and $\lambda_0=300$ nm. As shorthand, we refer to this photoprocess under the assumption that $\lambda_0=$Y nm by CuCN3-Y. Figure~\ref{fig:actspec} presents these action spectra.  

Action spectra typically encode information about relative, not absolute, reaction rates. Consequently, they are generally arbitrarily normalized to 1 at some wavelength (see, e.g., \citealt{Cockell1999} and \citealt{Rugheimer2015}). We normalize these spectra to 1 at 190 nm.

\section{Results\label{sec:results}}
\subsection{Clear-Sky H$_2$O-CO$_2$ Atmospheres}
\begin{figure}[H]
\centering
\includegraphics[width=10 cm, angle=0]{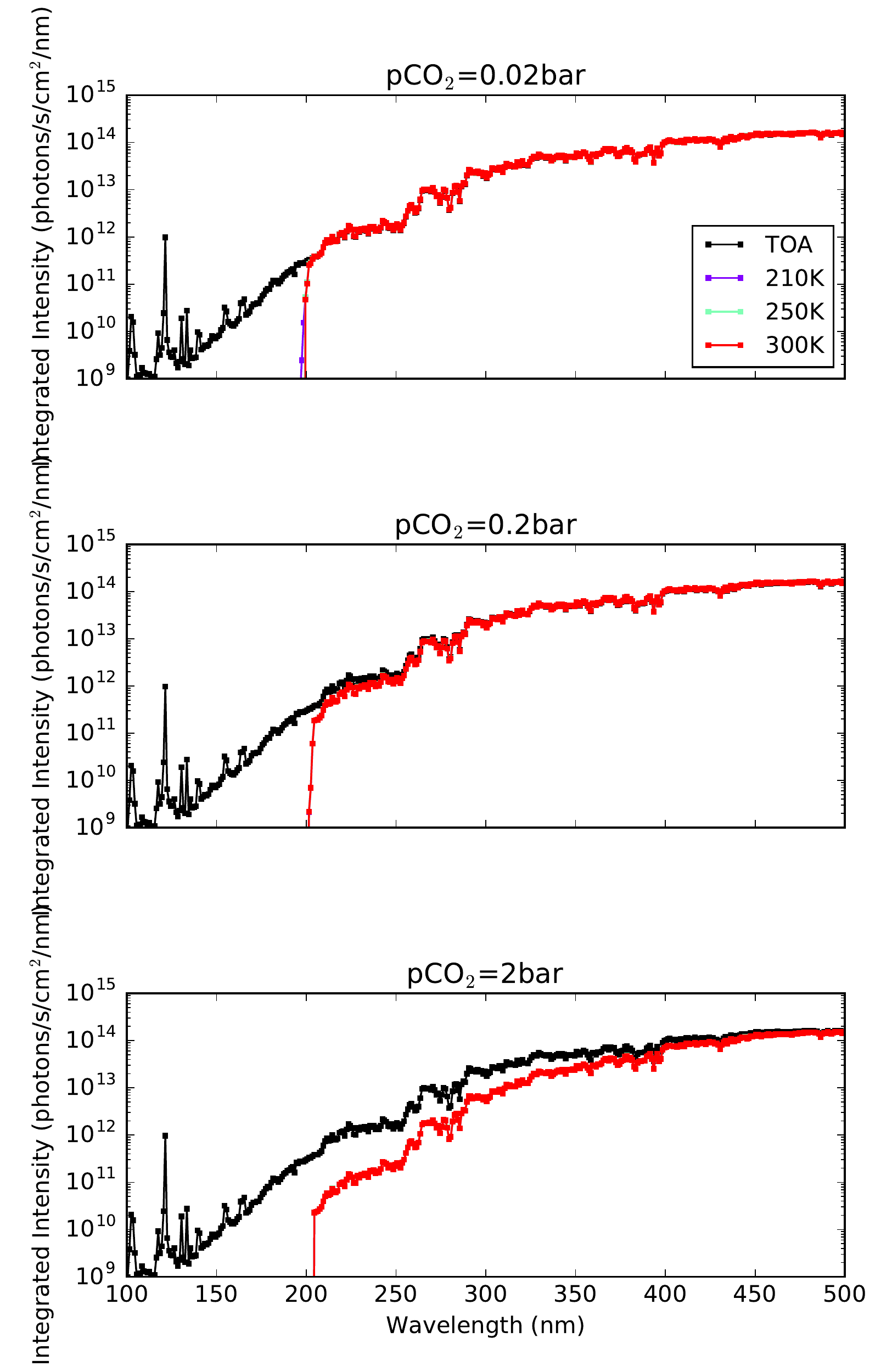}
\caption{Surface radiance as a function of wavelength for varying pCO$_2$ and $T_0$, for SZA=0 and albedo corresponding to desert. Also plotted for scale is the TOA solar flux. The surface radiance is insensitive to atmospheric and surface temperature. \label{fig:tpdep}}
\end{figure}

We evaluated the UV surface radiance for a range of (pCO$_2$ , $T_0$) for pure H$_2$O-CO$_2$ atmospheres in the clear-sky case (no clouds, dust or other particulates). We considered pCO$_2=0.02-2$ bar, corresponding to the range of surface pressure for which \citealt{Wordsworth2013} reported at least transient local temperatures above 273 K, and $T_0=210-300$ K. We took an SZA of $0$, corresponding to noon at equatorial latitudes. We took the surface albedo to correspond to desert (diffuse albedo of 0.22); we adopted this albedo because 1) young Mars is thought to have been dry and desertlike in conventional climate models, and 2) the desert diffuse albedo corresponds roughly to modern Mars's surface albedo (e.g., \citealt{Kasting1991}). We note that variations in surface albedo and SZA can drive variations in the spectral surface radiance of up to a factor of $\sim20$, and overall variations in prebiotically-relevant reaction rates of a factor of $\gtrsim10$ \citep{Ranjan2016b}. These surface radiances are shown in Figure~\ref{fig:tpdep}. 

In the scattering regime ($\lambda>204$ nm), our surface radiances fall off only slowly with pCO$_2$. This is a consequence of random walk statistics in the context of multiple scattering: transmission through purely scattering media go as $1/\tau$ (see, e.g., \citealt{Bohren1987} for a discussion with application to clouds). This result stands in contrast to the calculations of \citet{Cnossen2007} and \citet{Ronto2003}, who ignore multiple scattering in their radiative transfer treatments, and illustrates the importance of self-consistently including this phenomenon when considering dense, highly scattering atmospheres.

We find our surface radiance calculations to be insensitive to $T_0$. This is because 1) the total atmospheric column is set by $P_0$ and independent of $T_0$, 2) the rapid increase of CO$_2$ cross-sections for $\lambda<204$ nm, which means that the the atmosphere rapidly becomes optically thick in the UV, and 3) increased water vapor abundance with increasing $T_0$ does not drive an increase in opacity because water vapor absorption is degenerate with CO$_2$ absorption in the UV \citep{Ranjan2016b}.

We considered the hypothesis that including the effect of variations in CO$_2$ cross-section with temperature might impact our results. We followed the approach of \citet{Hu2012} in interpolating between cold ($\sim195$ K) and room-temperature datasets for CO$_2$ absorption to estimate the effects of temperature dependence on CO$_2$ cross-section. We used the dataset of \citet{Stark2007} from 106.5-118.7 nm, \citet{Yoshino1996} from 118.7-163 nm, and \citet{Parkinson2003} from 163-192.5 nm. We did not find cold-temperature cross-sections for CO$_2$ at longer wavelengths. We found our results were not altered by including temperature-dependence of CO$_2$ cross-sections for pCO$_2=0.02-2$ bar, because the CO$_2$ UV absorption is already saturated by 192.5 nm, where our temperature dependence kicks in. We considered lower values of pCO$_2=2\times10^{-3}-2\times10^{-5}$ bar (below PAL), where the CO$_2$ absorption does not saturate until wavelengths shorter than 192.5 nm.  Even in the low pCO$_2$ case, including temperature-dependence only changed the onsite of CO$_2$ absorption saturation by $1-2$ nm. We attribute this to the rapidity of the rise in CO$_2$ absorption cross section with decreasing wavelength for $\lambda \lesssim 204$ nm. We conclude that even including the effects of temperature on CO$_2$ UV cross-section, the UV surface fluence is insensitive to $T_0$. We consequently elected to ignore the temperature dependence of CO$_2$ cross-sections in the remainder of this study.

\begin{figure}[H]
\centering
\includegraphics[width=10 cm, angle=0]{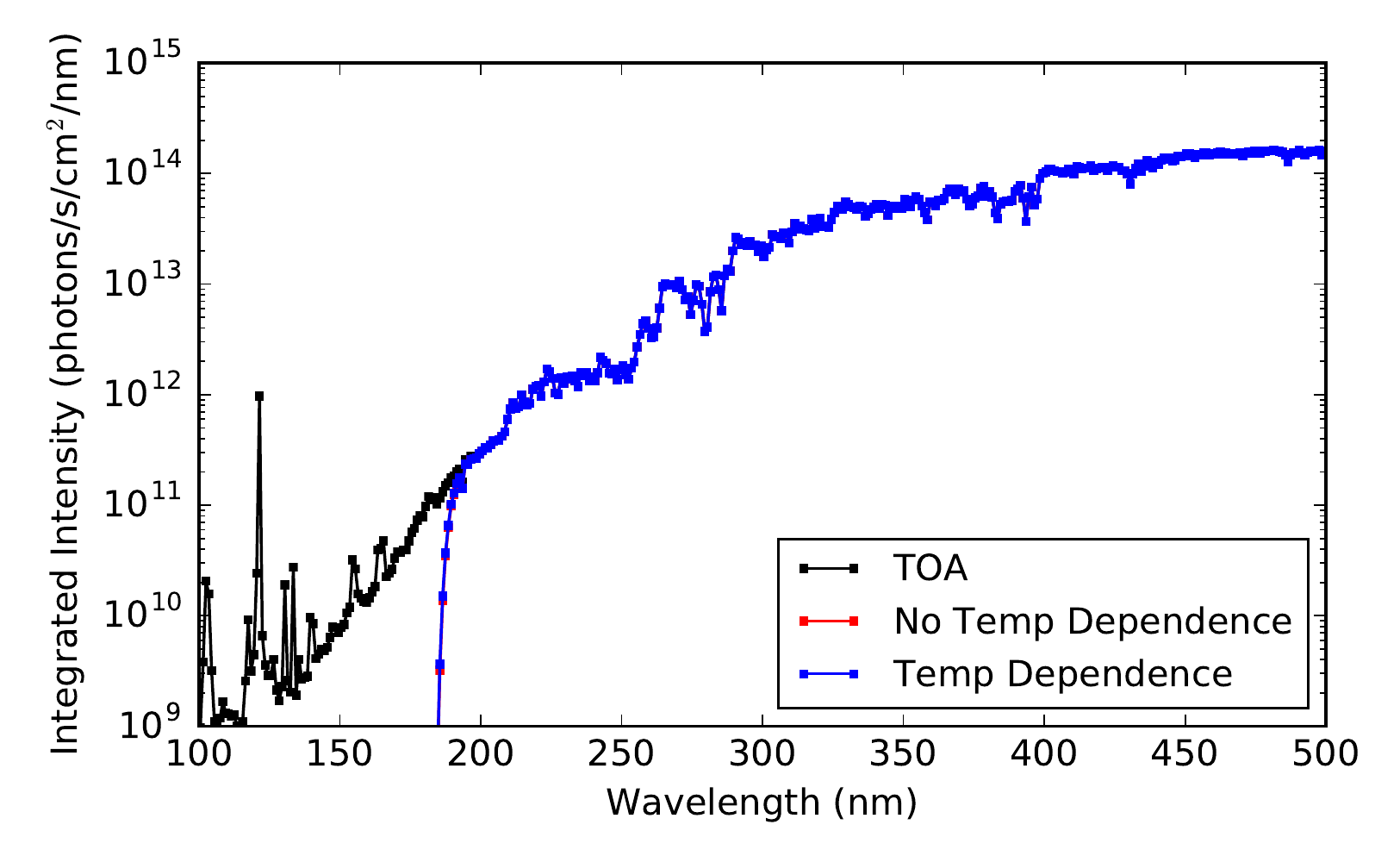}
\caption{Surface radiance as a function of wavelength for pCO$_2=2\times10^{-5}$ bar , $T_0=200$K, SZA=0,  $A$ corresponding to desert, with and without temperature dependence of the CO$_2$ cross-sections included. Also plotted for scale is the TOA solar flux. The surface radiance is insensitive to inclusion of temperature dependence of the CO$_2$ cross-sections, even for very low pCO$_2$. \label{fig:td}}
\end{figure}

Figure~\ref{fig:td} presents the surface radiances in the pCO$_2=2\times10^{-5}$ bar cases, calculated for SZA=0, $A$ corresponding to desert, and $T_0=T_{eq}\approx200$K, with and without CO$_2$ cross-sections included. Such low atmospheric pressures have been suggested based on atmospheric escape arguments \citep{Tian2009}, followed by a buildup of the atmosphere after escape rates subsided with shortwave solar output. In such a case, aqueous prebiotic chemistry could only have proceeded in environments kept warm by non-climatological means, e.g. geothermal reservoirs. Even for such low pCO$_2$, EUV fluence shortward of 185 nm is shielded out by atmospheric CO$_2$.

\subsection{Effect of CO$_2$ and H$_2$O Clouds}
\begin{figure}[H]
\centering
\includegraphics[width=10 cm, angle=0]{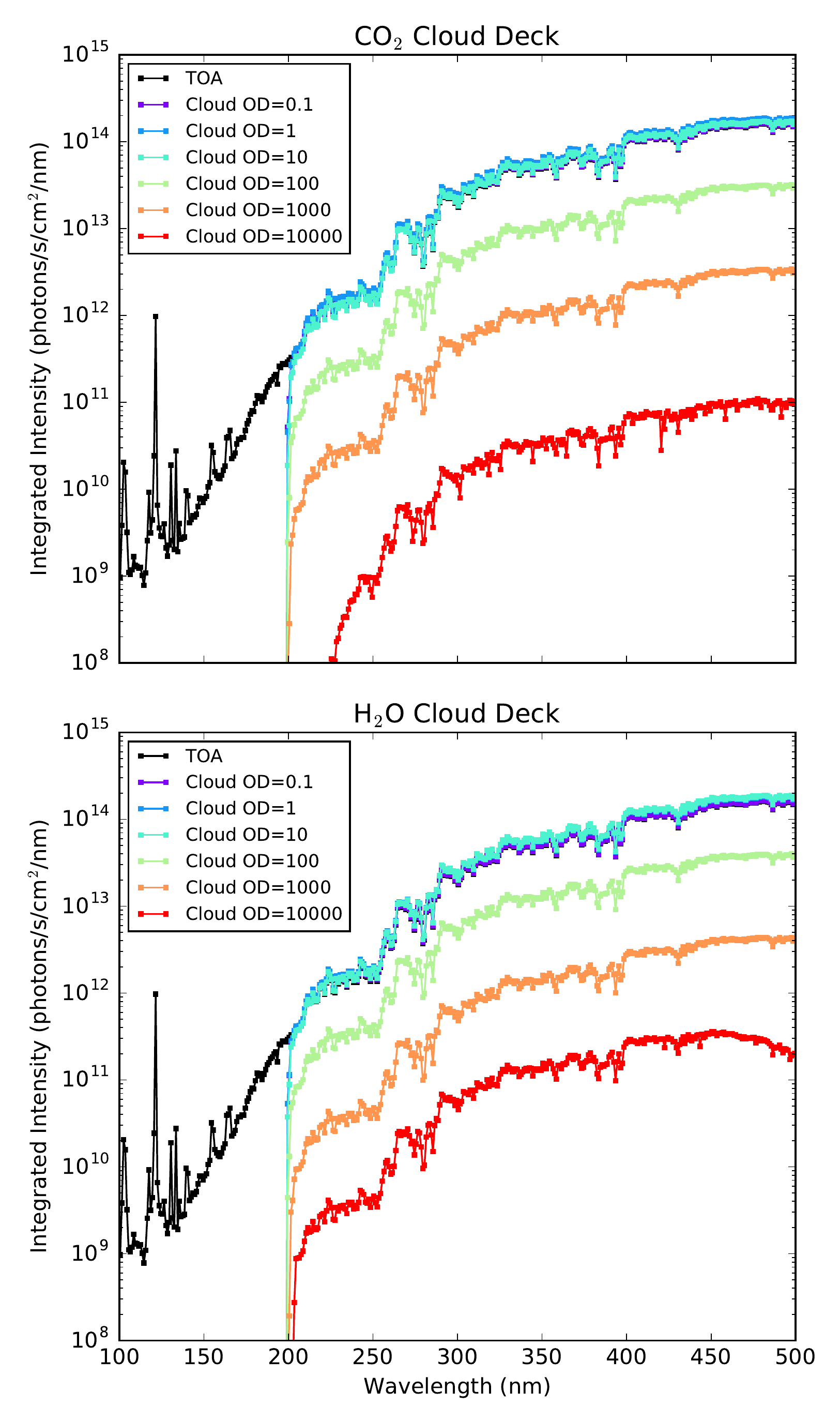}
\caption{Surface radiance as a function of wavelength for pCO$_2=0.02$ bar, $T_0=250$K, SZA=0,  $A$ corresponding to desert, and CO$_2$ and H$_2$O cloud decks of varying thicknesses inserted from 20-21 and 3-4 km respectively. Also plotted for scale is the TOA solar flux. \label{fig:results_clouds}}
\end{figure}

We considered the effect of CO$_2$ and H$_2$O clouds in a CO$_2$-H$_2$O atmosphere. Such clouds have been detected on modern Mars (see, e.g., \citealt{Vincendon2011}), and GCM results suggest they should have been present on early Mars as well\citep{Wordsworth2013b}. Figure~\ref{fig:results_clouds} presents the UV surface fluence for a 0.02-bar CO$_2$-H$_2$O atmosphere with H$_2$O and CO$_2$ cloud decks of varying optical depths emplaced in the atmosphere. This low surface pressure is chosen in order to isolate the effects of the clouds as opposed to atmospheric CO$_2$. The surface albedo corresponds to desert, and SZA=0. The optical depths are specified at 500 nm. The H$_2$O and CO$_2$ cloud decks are emplaced from 3-4 km and 20-21 km of altitude, respectively, corresponding approximately to the altitudes of peak cloud formation identified in \citet{Wordsworth2013}. For both types of clouds, we varied the cloud deck altitudes between 0.5-60.5 km, and found the surface fluence to be insensitive to the cloud deck altitude. We also experimented with partitioning the clouds into two decks, and found the surface fluence to be insensitive to the partition.

CO$_2$ and H$_2$O ice clouds have similar impact on the surface fluence. This is because both types of ice have similar optical parameters for $195<\lambda<500$ nm and $r_{eff}\geq1 \mu$m. The $\lambda<195$ nm regime, where they have different optical parameters, is shielded out by gaseous CO$_2$ absorption. 

CO$_2$ and H$_2$O ice particles are pure scatterers for $195<\lambda<500$ nm and $r_{eff}\geq1 \mu$m. Consequently, it is unsurprising that surface fluence falls off only slowly with increasing optical depth. In fact, the transmission of a purely scattering cloud layer varies as $\sim 1/\tau^*$, where $\tau^*$ is the delta-scaled optical depth of the cloud layer \citep{Bohren1987}. This is a consequence of the random-walk nature of radiative transfer in the optically-thick purely-scattering limit. 

The particle size $r_{eff}$ can have an impact on surface radiance. For $r_{eff}=1 \mu$m, the surface fluence at 500 nm is $\sim 40\%$ lower compared to $r_{eff}=100 \mu$m ( $A$ corresponding to desert, SZA=0). This is because as $r_{eff}$ decreases in this regime, so does $g$, meaning that the rescaled optical depth in the delta-scaling formalism $\tau^{*}=\tau\times(1-g^2)$ is higher for small particles than large particles. 

Suppressing surface radiance to $10\%$ or less of TOA flux requires cloud optical depths of $\gtrsim100$. This is comparable to the optical depth of a terrestrial thunderstorm \citep{Mayer1998}. We can compute the mass column of ice particle required to achieve this optical depth by $$u=(\frac{\tau}{\pi r^2 Q_{ext}}) (\frac{4\pi}{3}r^3 \rho),$$ where $\rho$ is the mass density of the ice, $r$ is the ice particle radius, $\tau$ is the cloud optical depth, and $Q_{ext}$ is the extinction efficiency. For CO$_2$ ice, $\rho=1.5$ g/cm$^3$ \footnote{\url{http://terpconnect.umd.edu/~choi/MSDS/Airgas/CARBON\%20DIOXIDE.pdf}}, $r=10 \mu$m, and approximating $Q_{ext}=2$, we find $\tau=100$ corresponds to $u=1$ kg/m$^2$. \citet{Wordsworth2013b} find in their 3D simulation CO$_2$ ice columns of up to $0.6$ kg/m$^2$ in patches, suggesting CO$_2$ clouds may have significantly affected the surface UV environment. For water ice, taking $\rho=0.92$ g/cm$^3$ \citep{CRC90}, $r=10 \mu$m, and approximating $Q_{ext}\approx2$, we find $\tau=100$ corresponds to $u=6\times10^{-1}$ kg/m$^2$. By contrast, \citet{Wordsworth2013b} finds expected ice columns of $\lesssim 2\times10^{-3}$ kg/m$^2$. Consequently, barring a mechanism which can increase the H$_2$O cloud levels above that considered by \citet{Wordsworth2013b}, H$_2$O ice clouds by themselves are unlikely to significantly alter the surface radiance environment on early Mars.

\subsection{Effect of Elevated Levels of Volcanogenic Gases}
So far, we have considered pure CO$_2$-H$_2$O atmospheres. However, other gases may have been present in the early Martian atmosphere. In particular, Mars at $\sim3.9$ Ga was characterized by volcanic activity, which emplaced features like the Tharsis igneous province \citep{Halevy2014}. Such volcanism could have injected elevated levels of volcanogenic gases like SO$_2$ and H$_2$S into the atmosphere. SO$_2$ and H$_2$S are also strong and broad UV absorbers, and at elevated levels they can completely reshape the surface UV environment \citep{Ranjan2016b}. Consequently, we consider the impact of elevated levels of SO$_2$ and H$_2$S on the surface UV environment.

\begin{figure}[H]
\centering
\includegraphics[width=10 cm, angle=0]{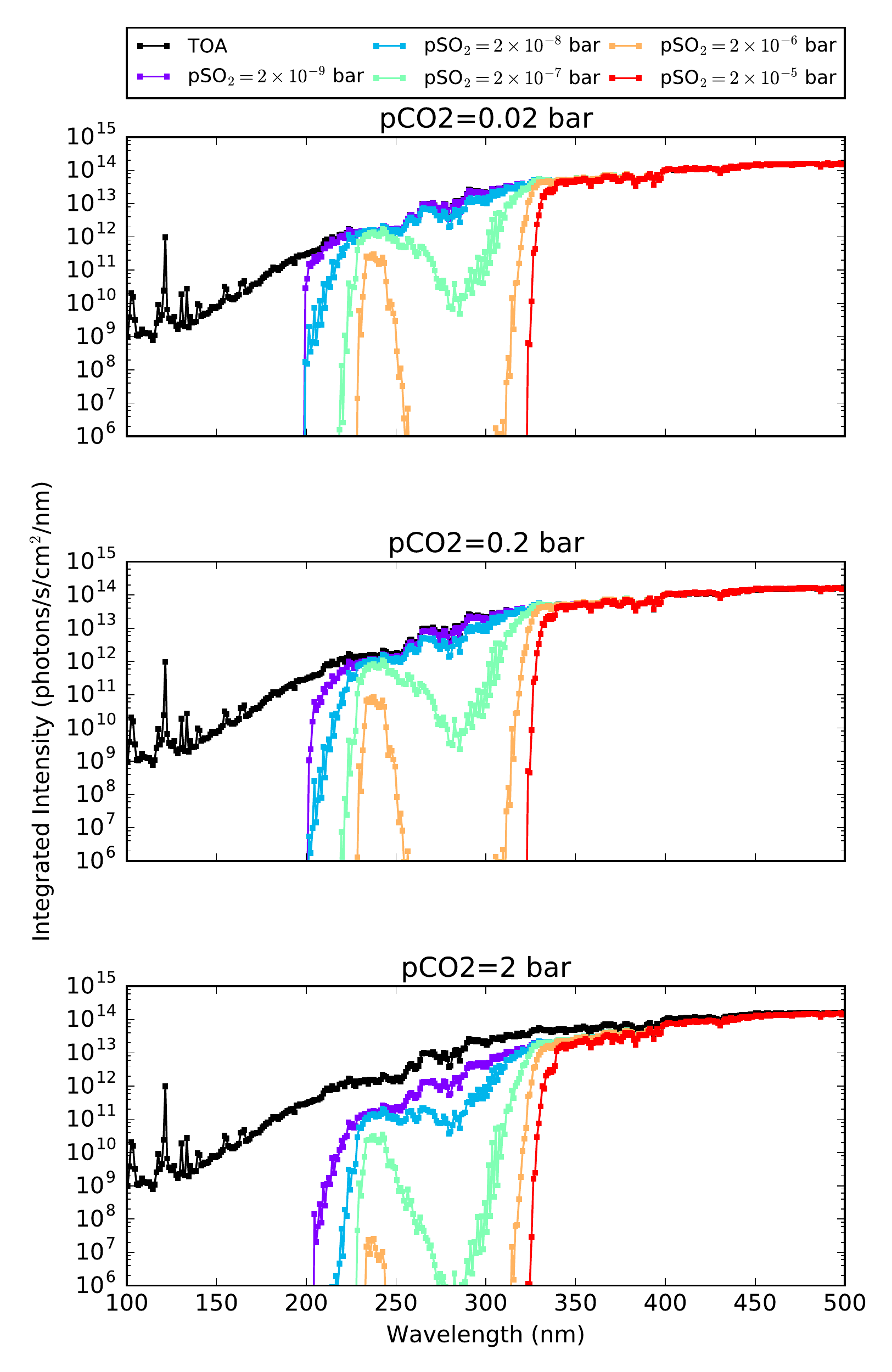}
\caption{Surface radiance as a function of wavelength for $T_0=250$K, SZA=0, $A$ corresponding to desert, pCO$_2=0.02-2$ bar, and pSO$_2=2\times10^{-9}-2\times10^{-5}$ bar. Also plotted for scale is the TOA solar flux. \label{fig:results_pso2_pco2}}
\end{figure}

We consider SO$_2$ levels up to $2\times10^{-5}$ bar. For scale, \citet{Halevy2014} compute that an SO$_2$ level of $1\times10^{-5}$ bar in a 1 bar CO$_2$-dominated atmosphere requires volcanic outgassing at $100\times$ the current terrestrial outgassing. Figure~\ref{fig:results_pso2_pco2} presents the surface radiance for varying pSO$_2$ and pCO$_2=0.02-2$ bar (SZA=0, $A$ corresponding to desert, and $T_0=250$ K). 

Note that for the same pSO$_2$, the surface radiance varies as a function of pCO$_2$. This is because at high pCO$_2$, the scattering optical depth of the atmosphere exceeds unity. In such a regime, the impact of trace absorbers is amplified due to higher effective path length caused by multiple scattering events \citep{Bohren1987}. This effect has been seen in studies of UV transmittance through thick clouds on Earth as well \citep{Mayer1998}. Consequently, the surface UV environment is a feature of both pCO$_2$ and pSO$_2$. For pSO$_2\geq\times10^{-5}$ bar, UV fluence $<330$ nm is strongly suppressed.

\begin{figure}[H]
\centering
\includegraphics[width=10 cm, angle=0]{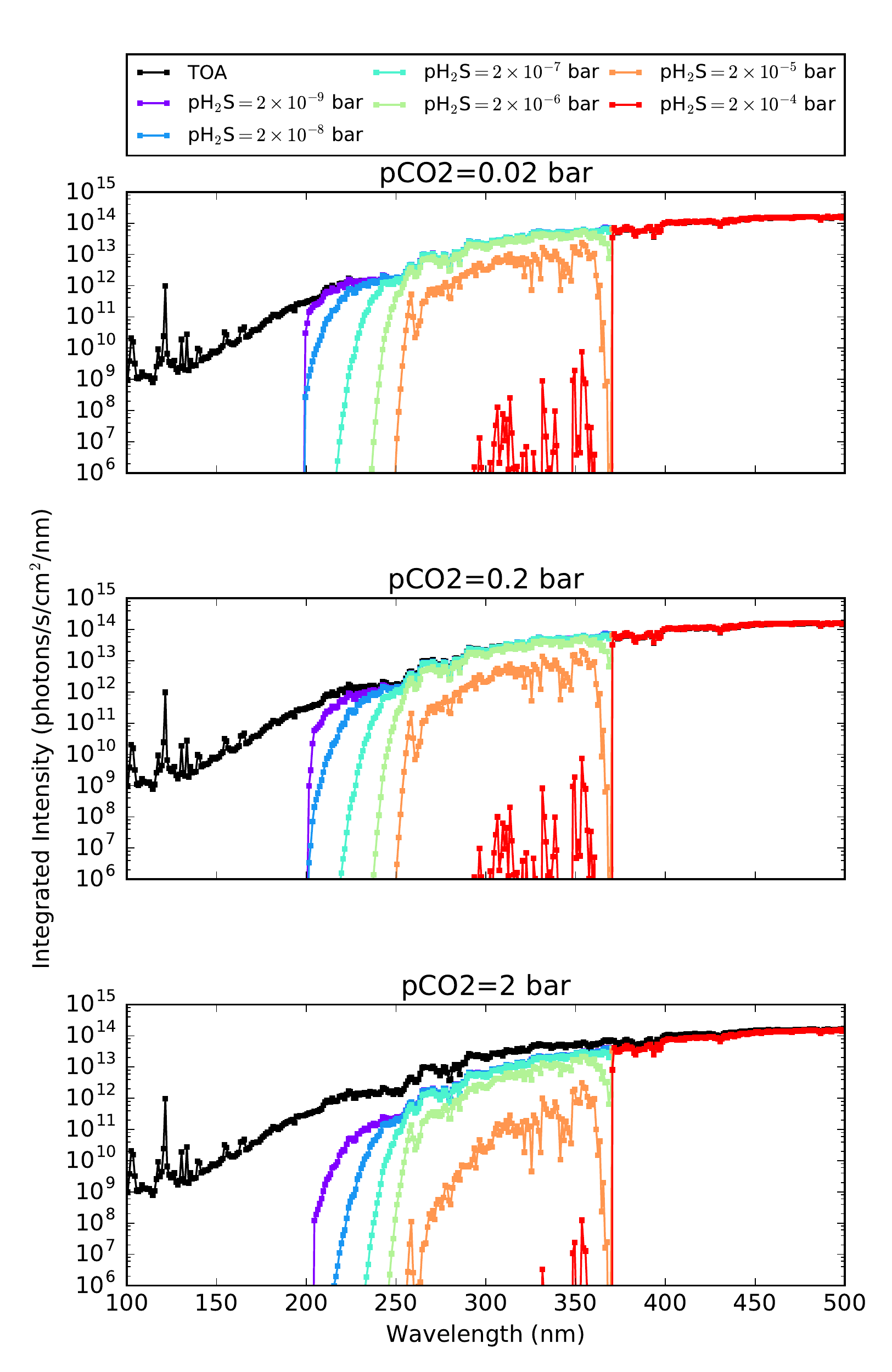}
\caption{Surface radiance as a function of wavelength for $T_0=250$K, SZA=0, $A$ corresponding to desert, pCO$_2=0.02-2$ bar, and pH$_2$S$=2\times10^{-9}-2\times10^{-4}$ bar. Also plotted for scale is the TOA solar flux.\label{fig:results_ph2s_pco2}}
\end{figure}

H$_2$S is also a major volcanogenic gas that may have been emitted at rates greater than or equal to SO$_2$ on young Mars, based on studies of the oxidation state of Martian basalts \citep{Herd2002, Halevy2007}. While \citet{Halevy2014} do not calculate the H$_2$S abundance as a function of volcanic outgassing, the calculations of \citet{Hu2013} (T$_0$=288K, modern solar irradiance) suggest that pH$_2$S$>$pSO$_2$ in a CO$_2$ dominated atmosphere. Consequently, we also consider the impact of elevated levels of H$_2$S, up to pH$_2$S=$2\times10^{-4}$ bar in atmospheres with pCO$_2=0.02-2$ bar. Figure~\ref{fig:results_ph2s_pco2} presents the surface radiance calculated over this range of atmospheres (SZA=0, $A$ corresponding to desert, and $T_0=250$ K). For  pH$_2S\geq\times10^{-4}$ bar, UV fluence $<370$ nm is strongly suppressed.

\begin{figure}[H]
\centering
\includegraphics[width=10 cm, angle=0]{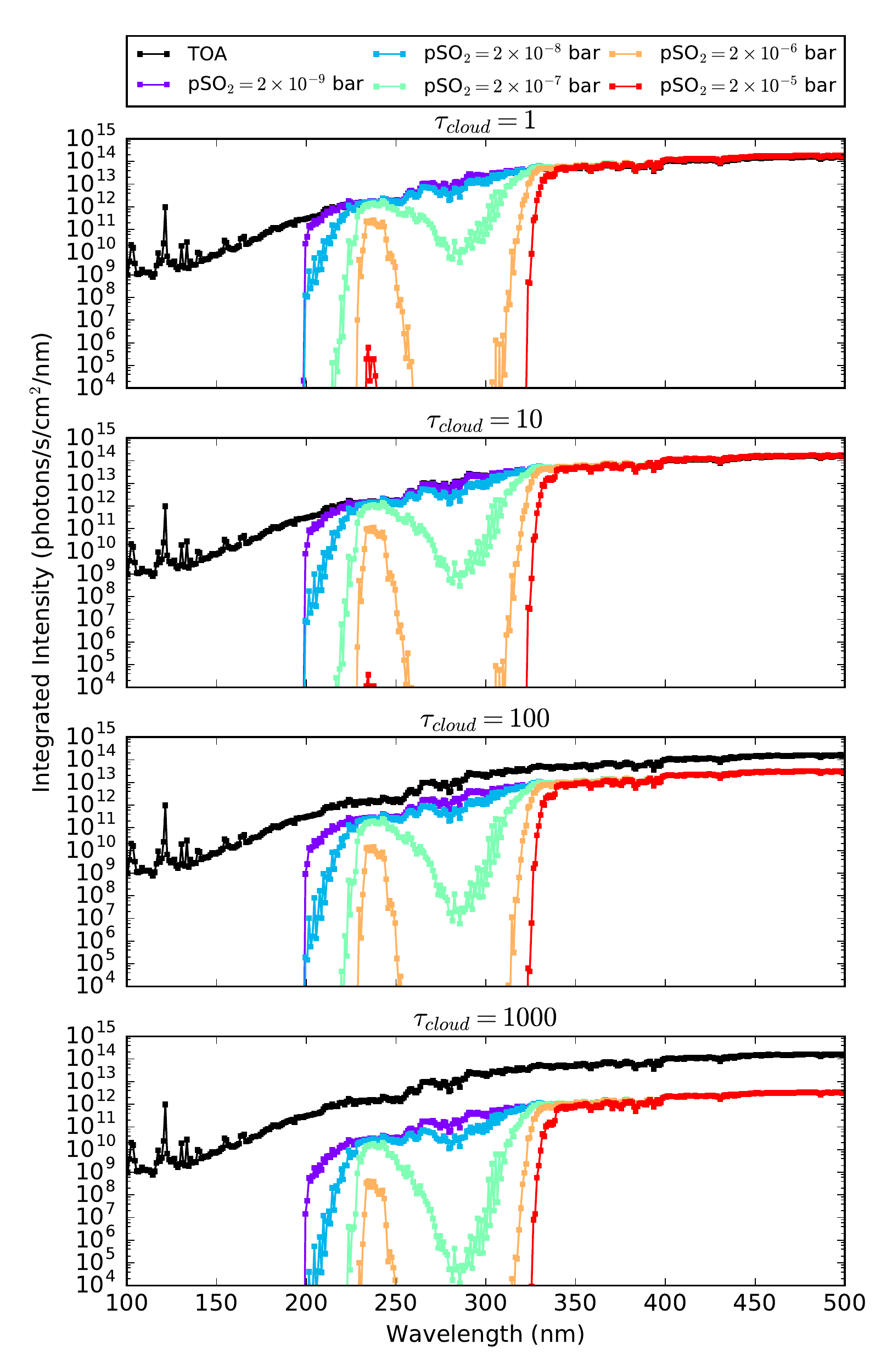}
\caption{Surface radiance as a function of wavelength for $T_0=250$K, SZA=0, $A$ corresponding to desert, pCO$_2=0.02$ bar, pSO$_2=2\times10^{-9}-2\times10^{-5}$ bar, and CO$_2$ cloud decks of varying optical thickness emplaced from 20-21 km altitude. Also plotted for scale is the TOA solar flux. \label{fig:results_pso2_clouds}}
\end{figure}

\begin{figure}[H]
\centering
\includegraphics[width=10 cm, angle=0]{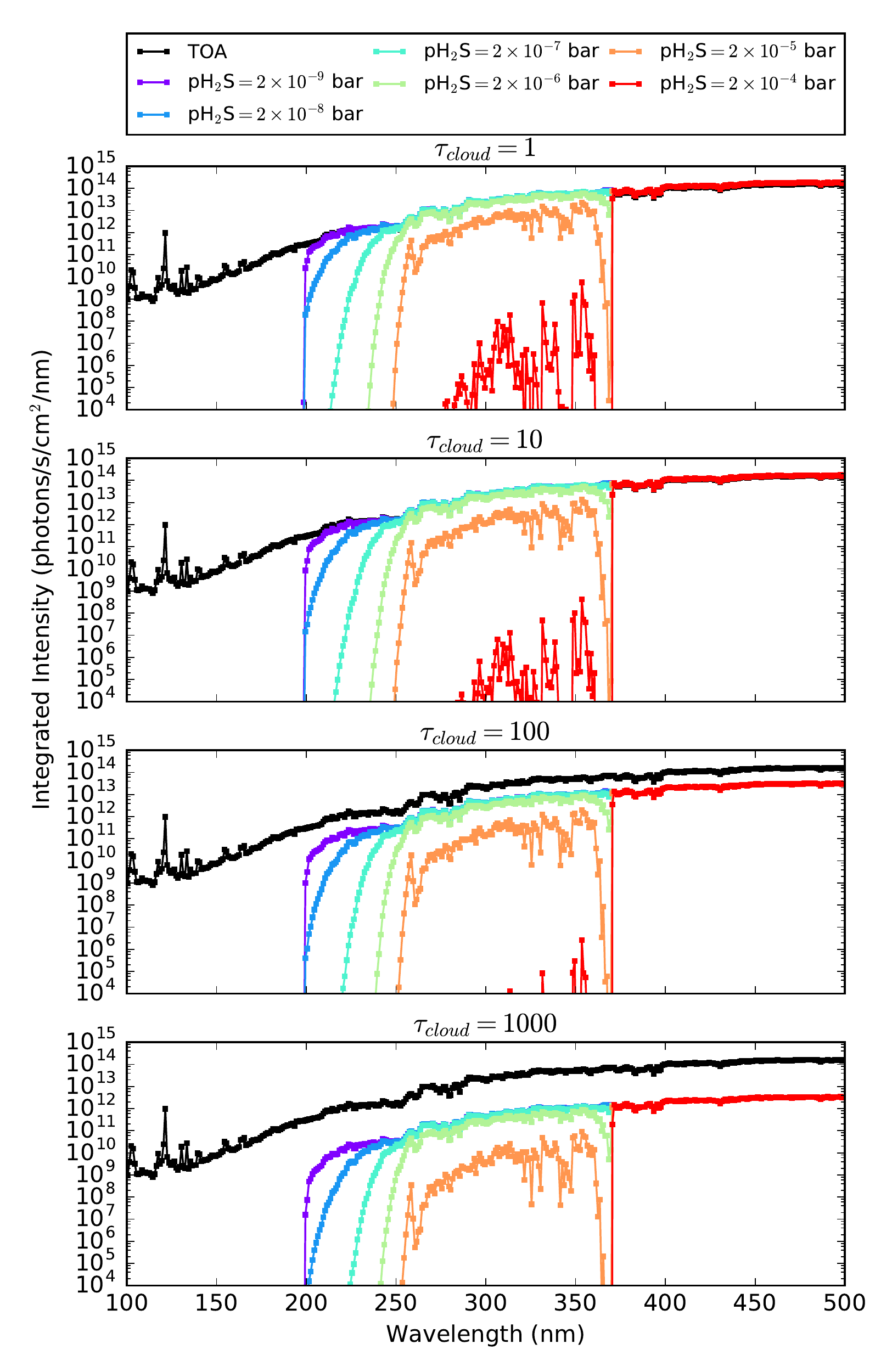}
\caption{Surface radiance as a function of wavelength for $T_0=250$K, SZA=0, $A$ corresponding to desert, pCO$_2=0.02$ bar, pH$_2$S$=2\times10^{-9}-2\times10^{-4}$ bar, and CO$_2$ cloud decks of varying optical thickness emplaced from 20-21 km altitude. Also plotted for scale is the TOA solar flux.\label{fig:results_ph2s_clouds}}
\end{figure}

Scattering due to thick cloud decks can also enhance absorption by trace pSO$_2$ and pH$_2$S. For pSO$_2=2\times10^{-7}$ bar and pCO$_2$=0.02 bar\footnote{At pCO$_2$=0.02 bar, the gaseous scattering optical depth is less than unity for wavelengths longer than 204 nm, meaning that we can attribute this amplification primarily to the cloud deck (as opposed to gaseous scattering).} (SZA=0, $A$ corresponding to new snow (near 1), $T_0=250$ K, $r_{eff}=10 \mu$m), inclusion of CO$_2$ clouds with optical depth of $1000$ amplifies attenuation by a factor of $10$ at $236.5$ nm (optically thin in SO$_2$ absorption) and by a factor of $10^{5}$ at $281.5$ nm (optically thick in SO$_2$ absorption) compared to what one would calculate by multiplying the transmission from the SO$_2$ and cloud deck individually. This effect is weaker for low-albedo cases, because there are fewer passes of radiation through the atmosphere due to bouncing between the surface and cloud deck. This nonlinear variance in transmission due to multiple scattering effects illustrates the need to specify surface albedo, cloud thickness and absorber level when calculating surface radiances.  Figure~\ref{fig:results_pso2_clouds} and  Figure~\ref{fig:results_ph2s_clouds} present the surface radiance at the base of a pCO$_2$=0.02 bar atmosphere with varying levels of SO$_2$ and H$_2$S respectively, with CO$_2$ cloud decks of optical depths $1-1000$ emplaced from 20-21 km (SZA=0, $A$ corresponding to desert, $T_0=250$ K, $r_{eff}=10 \mu$m). 

\subsection{Effect of Dust}
The atmosphere of modern Mars is dusty, with typical dust optical depths varying from $\tau_d\sim0.2-2$ at solar wavelengths \citep{Smith2002, Lemmon2015}, with the higher values achieved during global dust storms. Mars's dryness contributes to its dustiness, through the availability of desiccated surface to supply dust and the lack of a hydrologic cycle to quickly scrub it from the atmosphere. If one assumes that Mars were similarly dry in the past, dust levels in the Martian atmosphere may have been significant. We may speculate that if the atmosphere were thicker, it could have hosted even more dust than the modern Martian atmosphere due to slower sedimentation times. However, detailed study of atmospheric dust dynamics, including analysis of how dust lofting scales with pCO$_2$, is required to constrain this possibility. Dust absorbs at UV wavelengths (see Figure~\ref{fig:mie}), so dust could play a role similar to volcanogenic gases in scrubbing UV radiative from the UV surface environment. 

We explored the impact of including dust in our calculation of surface UV fluence. We assumed that, similar to the modern Mars, the dust followed an exponential profile with scale height height $H_d=11$ km, similar to the atmospheric pressure scale height \citep{Hoekzema2010, Mishra2016}. For surface temperatures (and hence scale heights) comparable to modern Mars, this corresponds to the assumption by previous workers that the dust mixing ratio is constant in the lower atmosphere \citep{Forget1999}. Then, the dust optical depth across each atmospheric layer of width $\Delta z$ is $$C\exp{[-z/H_d]},$$ where $z$ is the altitude of the layer center. The parameter $$C=\tau_d(\exp{[\Delta z/(2H_d)]})(1-\exp{[-\Delta z/H_d]})$$, is chosen such that the column-integrated dust optical depth is $\tau_d$. 

\begin{figure}[H]
\centering
\includegraphics[width=10 cm, angle=0]{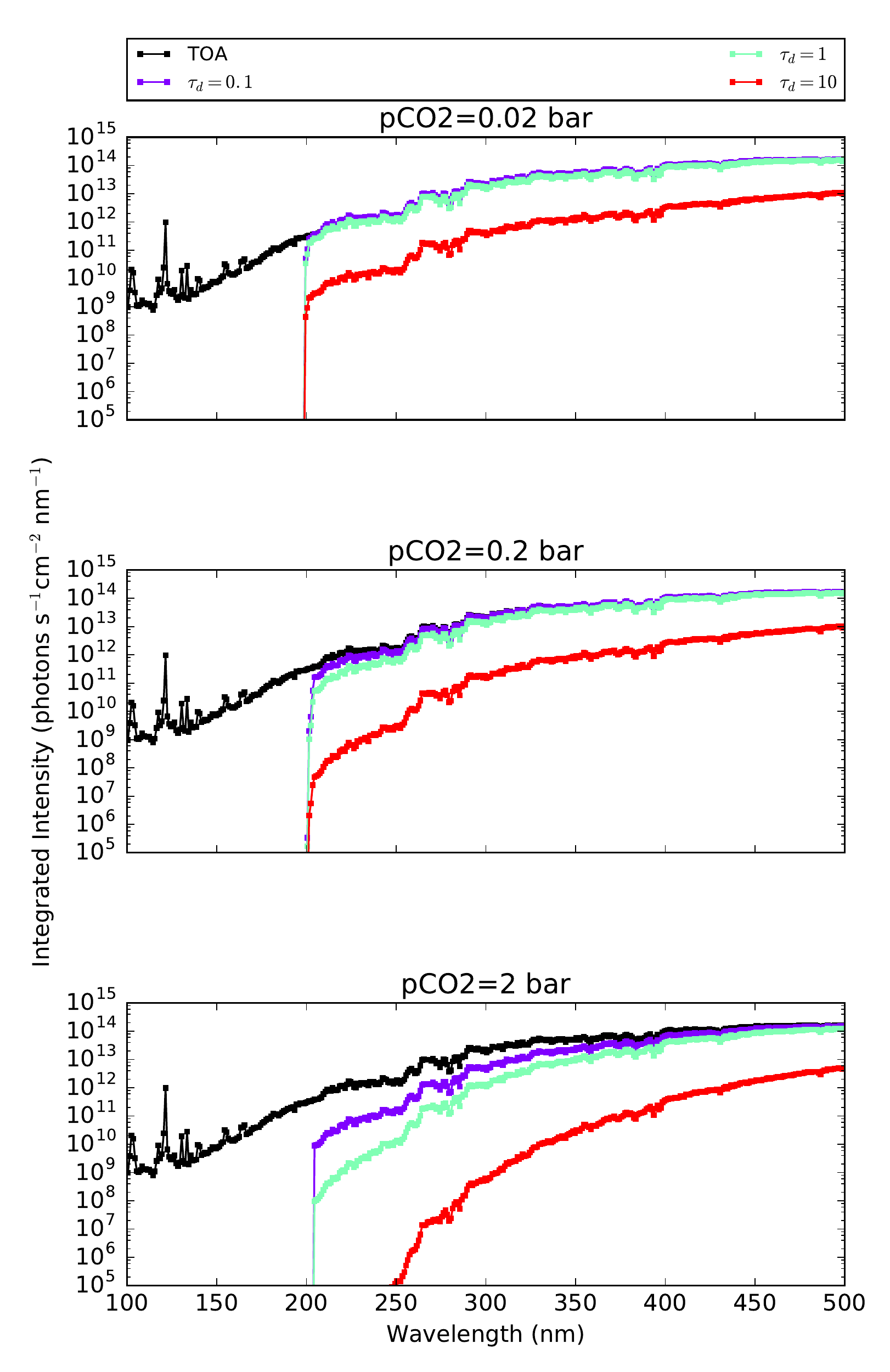}
\caption{Surface radiance as a function of wavelength for $T_0=250$K, SZA=0, $A$ corresponding to desert, pCO$_2=0.02-2$ bar, and $\tau_d=0.1-10$. Also plotted for scale is the TOA solar flux.\label{fig:results_dust_pco2}}
\end{figure}

Figure~\ref{fig:results_dust_pco2} presents the surface radiance for varying $\tau_d$ and pCO$_2=0.02-2$ bar (SZA=0, $A$ corresponding to desert, and $T_0=250$ K). As for SO$_2$ and H$_2$S, highly scattering atmospheres amplify the impact of trace absorbers. $\tau_d=1$ only marginally suppresses UV fluence for pCO$_2=0.02$ bar, but for pCO$_2=2$ bar shortwave fluence is suppressed due to enhanced Rayleigh scattering. In the absence of scattering amplification, $\tau_d\gtrsim10$ is required to strongly suppress UV fluence.

\begin{figure}[H]
\centering
\includegraphics[width=10 cm, angle=0]{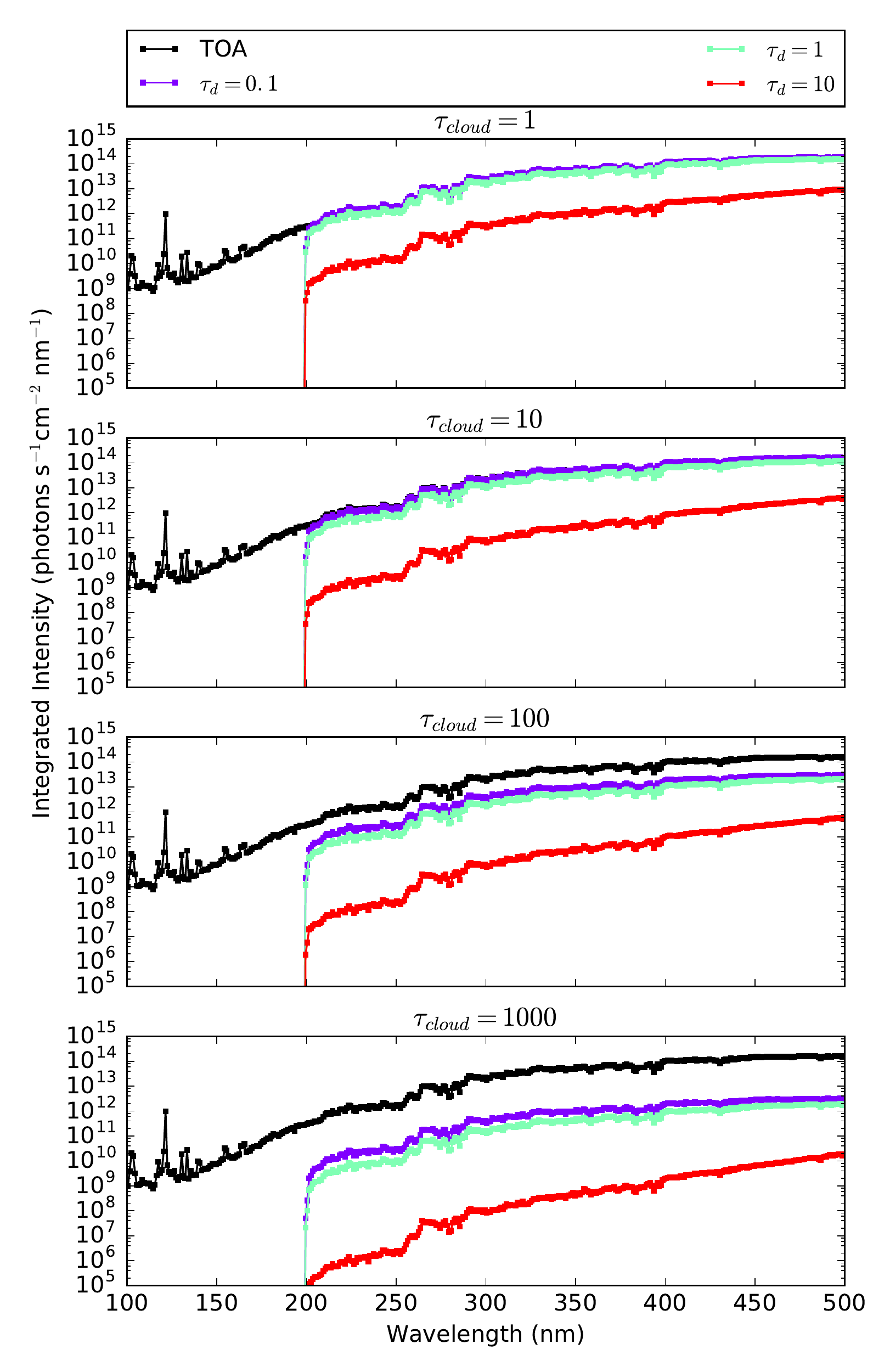}
\caption{Surface radiance as a function of wavelength for $T_0=250$K, SZA=0, $A$ corresponding to desert, pCO$_2=0.02$ bar, $\tau_d=0.1-10$, and CO$_2$ cloud decks of varying optical thickness emplaced from 20-21 km altitude. Also plotted for scale is the TOA solar flux.\label{fig:results_dust_clouds}}
\end{figure}

Figure~\ref{fig:results_dust_clouds} presents the surface radiance at the base of a pCO$_2$=0.02 bar atmosphere with varying $\tau_d$, with CO$_2$ cloud decks of optical depths $1-1000$ emplaced from 20-21 km (SZA=0, $A$ corresponding to desert, $T_0=250$ K, $r_{eff}=10 \mu$m). The impact of a cloud deck is less than the impact of increasing overall atmospheric pressure, because of the limited column of absorber contained within the cloud itself. If the cloud deck is extended, then the surface fluence is reduced (for the same total cloud and dust optical depth).

\section{Discussion\label{sec:discussion}}

\subsection{CO$_2$-H$_2$O Atmosphere\label{sec:co2h2o}}
Normative CO$_2$-H$_2$O climate models predict the steady-state early Martian climate to have been cold, with global mean temperatures below freezing \citep{Forget2013, Wordsworth2013b, Wordsworth2015}. In this scenario, aqueous prebiotic chemistry would proceed in meltwater pools, which could have occurred transiently during midday due to the diurnal cycle, during summer during a seasonal cycle, or a combination of both. Aqueous prebiotic chemistry could also have proceeded in geothermally heated pools, which may have been abundant during the more volcanically active Noachian. 

\begin{figure}[H]
\centering
\includegraphics[width=10 cm, angle=0]{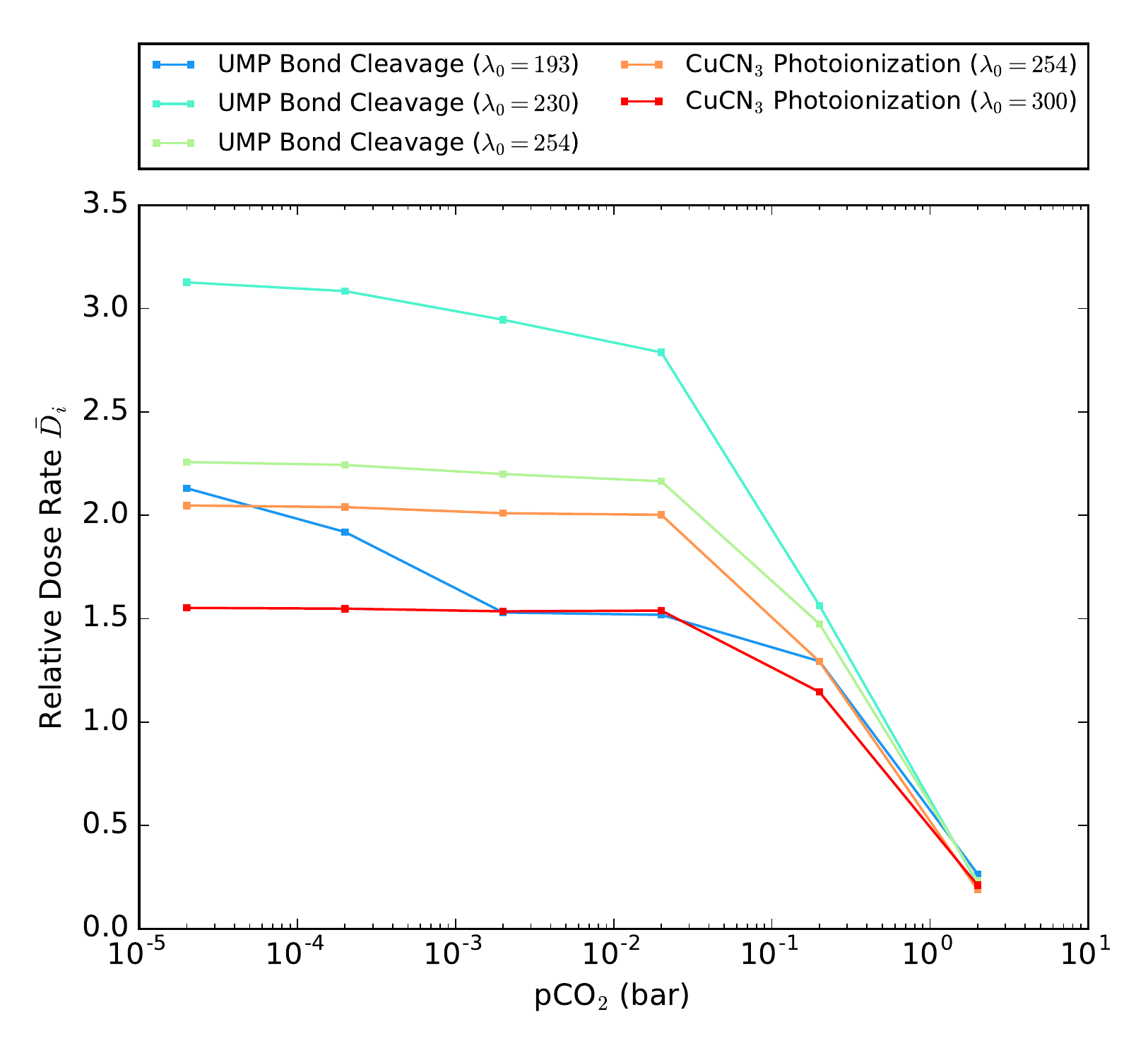}
\caption{UV dose rates $\overline{D}_i$ for clear-sky CO$_2$-H$_2$O atmospheres as a function of pCO$_2$ ($T_0=250$K, SZA=0, $A$ corresponding to desert). \label{fig:discussion_doses_pco2}}
\end{figure}

Figure~\ref{fig:discussion_doses_pco2} presents the dose rates $\overline{D}_i$ corresponding to irradiation of prebiotically relevant molecules through a clear-sky CO$_2$-H$_2$O atmosphere with SZA=0 and $A$ corresponding to desert. The dose rates $\overline{D}_i$ decline by only an order of magnitude across 5 orders of magnitude of pCO$_2\leq2$ bar, meaning that the prebiotic photochemistry dose rates are only weakly sensitive to pCO$_2$ across this range. $\overline{D}_i$ is within an order of magnitude of unity across this range, meaning that the Martian dose rates are comparable to the terrestrial rates. This is consistent with the results of \citet{Cockell2000}, who found that UV stress as measured by DNA damage were comparable for 3.5 Ga Earth and Mars. 

\begin{figure}[H]
\centering
\includegraphics[width=10 cm, angle=0]{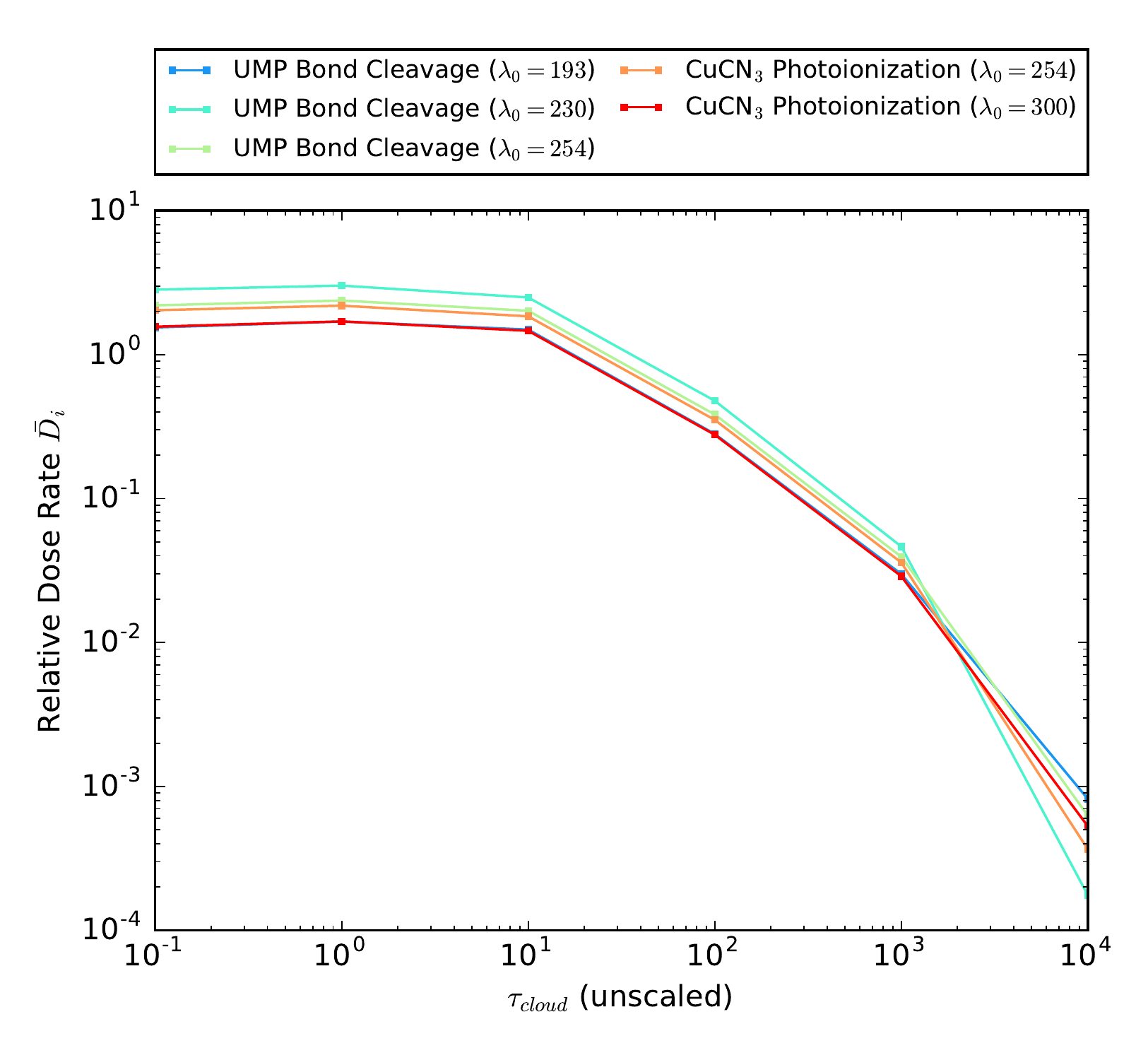}
\caption{UV dose rates $\overline{D}_i$ for CO$_2$-H$_2$O atmospheres with a CO$_2$ cloud deck emplaced from 20-21 km altitude, as a function of cloud deck optical depth ($T_0=250$K, pCO$_2$=0.02 bar, SZA=0, $A$ corresponding to desert). \label{fig:discussion_doses_co2clouds}}
\end{figure}

However, the sky was not necessarily clear during this epoch. In fact, formation of H$_2$O and CO$_2$ clouds are expected based on GCM studies \citep{Wordsworth2013b}. Thick (though patchy) CO$_2$ cloud decks in particular are expected for thick CO$_2$ atmospheres. Figure~\ref{fig:discussion_doses_co2clouds} presents the dose rates $\overline{D}_i$ as a function of CO$_2$ cloud optical depth for pCO$_2$=0.02 bar, SZA=0, and $A$ corresponding to desert. As we might expect from Figure~\ref{fig:results_clouds}, the dose rate drops off only modestly with $\tau_{cloud}$. $\tau_{cloud}\gtrsim1000$ is required to suppress dose rates by more than an order of magnitude. For $r_{eff}=10\mu m$, this corresponds to $20\times$ the maximum cloud column calculated by \citet{Wordsworth2013b}. Overall, the impact of clouds on their own on UV-sensitive photochemistry is expected to be modest. 

\begin{figure}[H]
\centering
\includegraphics[width=10 cm, angle=0]{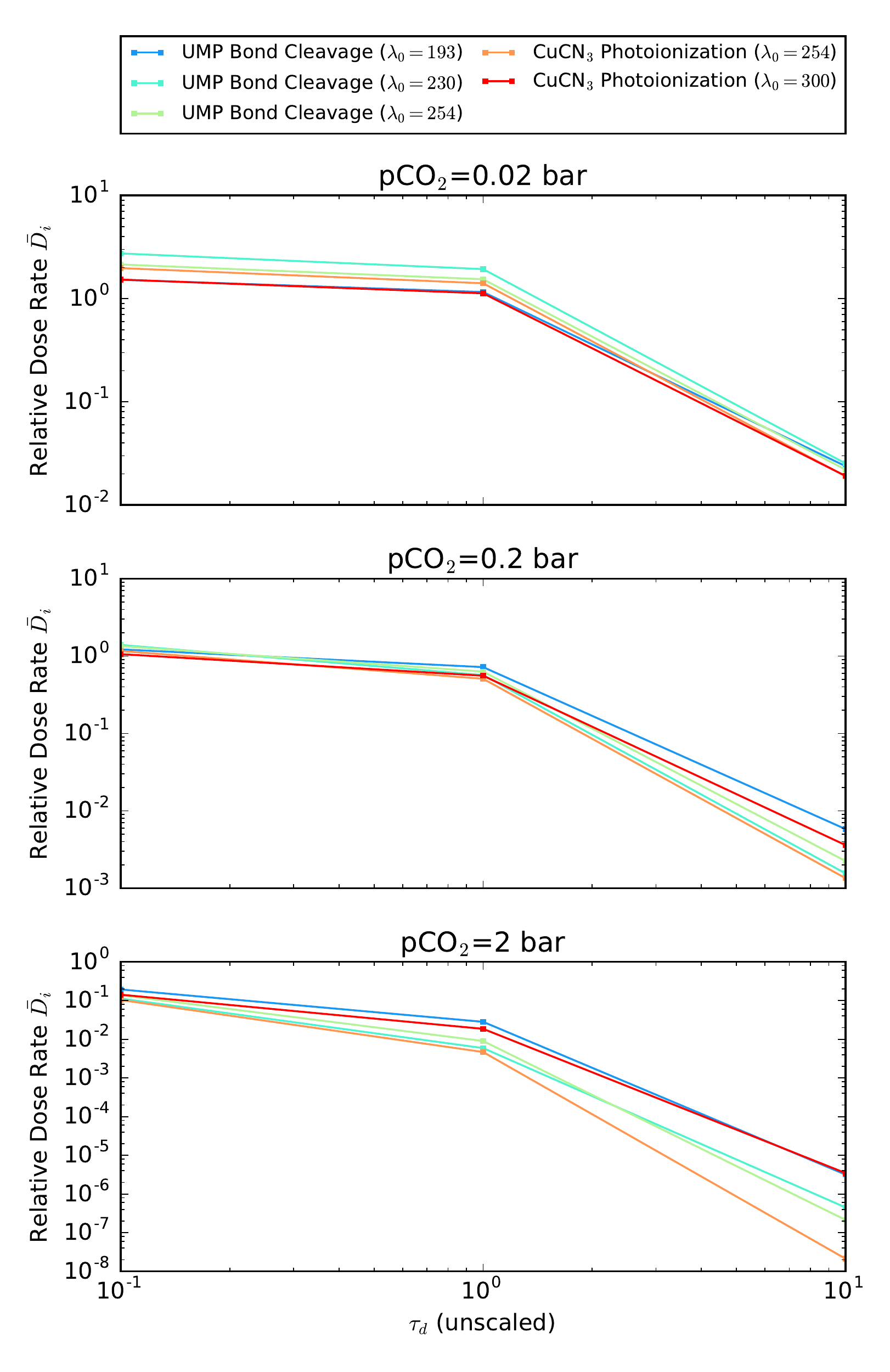}
\caption{UV dose rates $\overline{D}_i$ for dusty CO$_2$-H$_2$O atmospheres as a function of pCO$_2$ and $\tau_d$ ($T_0=250$K, SZA=0, $A$ corresponding to desert). \label{fig:discussion_doses_dust_pco2}}
\end{figure}

\begin{figure}[H]
\centering
\includegraphics[width=10 cm, angle=0]{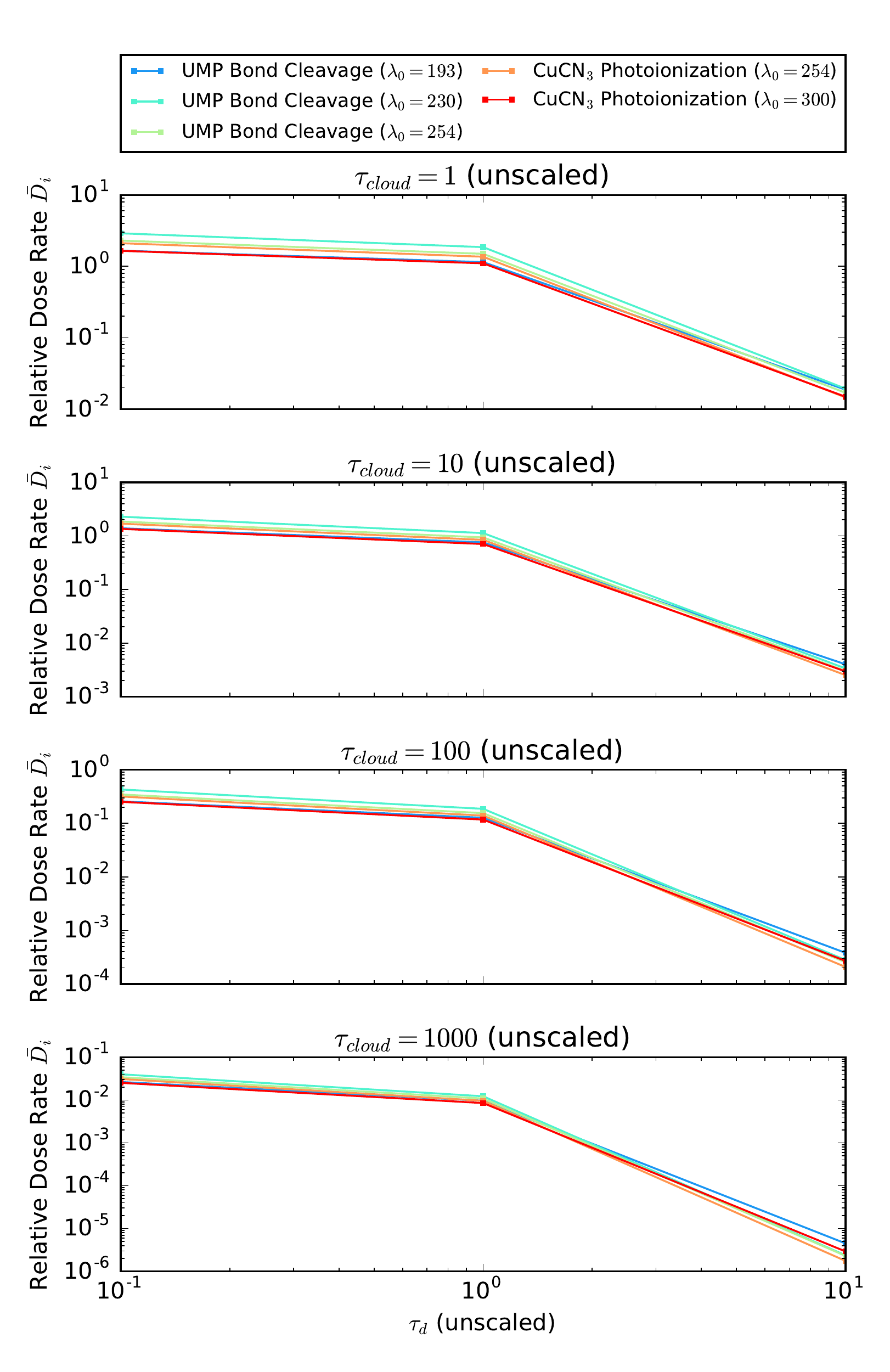}
\caption{UV dose rates $\overline{D}_i$ for dusty CO$_2$-H$_2$O atmospheres with a CO$_2$ cloud deck emplaced, as a function of $\tau_d$ and $\tau_{cloud}$ ($T_0=250$K, pCO$_2$=0.02, SZA=0, $A$ corresponding to desert). \label{fig:discussion_doses_dust_clouds}}
\end{figure}

If young Mars were indeed cold and dry, then one might expect it to have been dusty, as it is today. The modern Martian dust optical depth ranges from $\tau_d\sim0.2-2$, with higher values achieved during dust storms \citep{Smith2002, Lemmon2015}. Higher dust loadings may have been possible when the atmosphere was thicker. Figure~\ref{fig:discussion_doses_dust_pco2} presents the dose rates as a function of $\tau_d$ and pCO$_2$, for SZA=0 and $A$ corresponding to desert. Figure~\ref{fig:discussion_doses_dust_clouds} presents the dose rates as a function of $\tau_d$ and CO$_2$ cloud optical depth, for SZA=0 and $A$ corresponding to desert. 

Unlike CO$_2$ and H$_2$O ice and gas, dust absorbs across the UV waveband. Further, this absorption can be dramatically increased in highly scattering atmospheres due to enhanced effective path length due to multiple scattering effects. Consequently, dust can dramatically suppress UV fluence and hence photochemistry, especially for thick and/or cloudy atmospheres. For $\tau_d\geq10$, dose rates are suppressed by 2-8 orders of magnitude, depending on atmospheric pressure and cloud thickness. Dust of thickness $\tau_d=1$ provides minimal suppression for pCO$_2\leq$0.2 bar or $\tau_{cloud}<100$, but for pCO$_2\geq$2 bar or $\tau_{cloud}\geq1000$, can suppress dose rates by orders of magnitude. Dust of thickness $\tau_d\leq0.1$ does not significantly alter dose rates across the explored parameter space. Overall, dust can significantly alter dose rates if it is present at high levels (greater than that seen for modern Martian dust storms) or if it is present at levels comparable to the modern average, but embedded in a thick atmosphere or underlying thick clouds.

We considered the hypothesis that attenuation due to dust might differentially affect the stressor and eustressor pathways. That is, we considered the possibility that the eustressor reaction rates might fall off faster (or slower) than the stressor reaction rates with attenuation due to dust. Since the stressor pathway measures destruction of RNA monomers and the eustressor pathway measures a process important to the synthesis of key RNA precursors, this means that environments that favor the eustressor pathway over the stressor pathway are a more favorable venue for abiogenesis than the reverse, in the RNA world hypothesis. This argument assumes that these particular stressor and eustressor processes were important in the prebiotic world. They might not have been. However, we expect other photochemical stressor and eustressor processes to behave in generally similar ways to these processes. For example, we generally expect the quantum yield of prebiotic molecular destruction to decrease with increasing wavelength because of decreased photon energy compared to bond strength. Similarly, regardless of the solvated electron source (e.g. HS$^-$ as opposed to tricyanocuprate), we expect the quantum yield to go approximately as a step function in wavelength. We therefore suggest that results derived from these pathways may generalize to other processes, though a detailed comparison is required to rule on this hypothesis.

To assess the hypothesis that a dusty Mars might be less (or more) clement for abiogenesis than a non-dusty Mars as measured by our stressor (UMP-X) and eustressor (CuCN3-Y) pathways, we calculated $\overline{D}_{UMP-X}/\overline{D}_{CuCN3-Y}$. We calculated this quantity for pCO$_2=2$ bar (no clouds) and $\tau_{cloud}=1000$ (pCO$_2$=0.02 bar) for $\tau_d=0.1-10$. If these ratios rise with $\tau_d$, it means that the stressor pathway is relatively favored by dusty atmospheres; if they fall, it means that the stressor pathway is relatively disfavored by dusty atmospheres.

\begin{figure}[H]
\centering
\includegraphics[width=10 cm, angle=0]{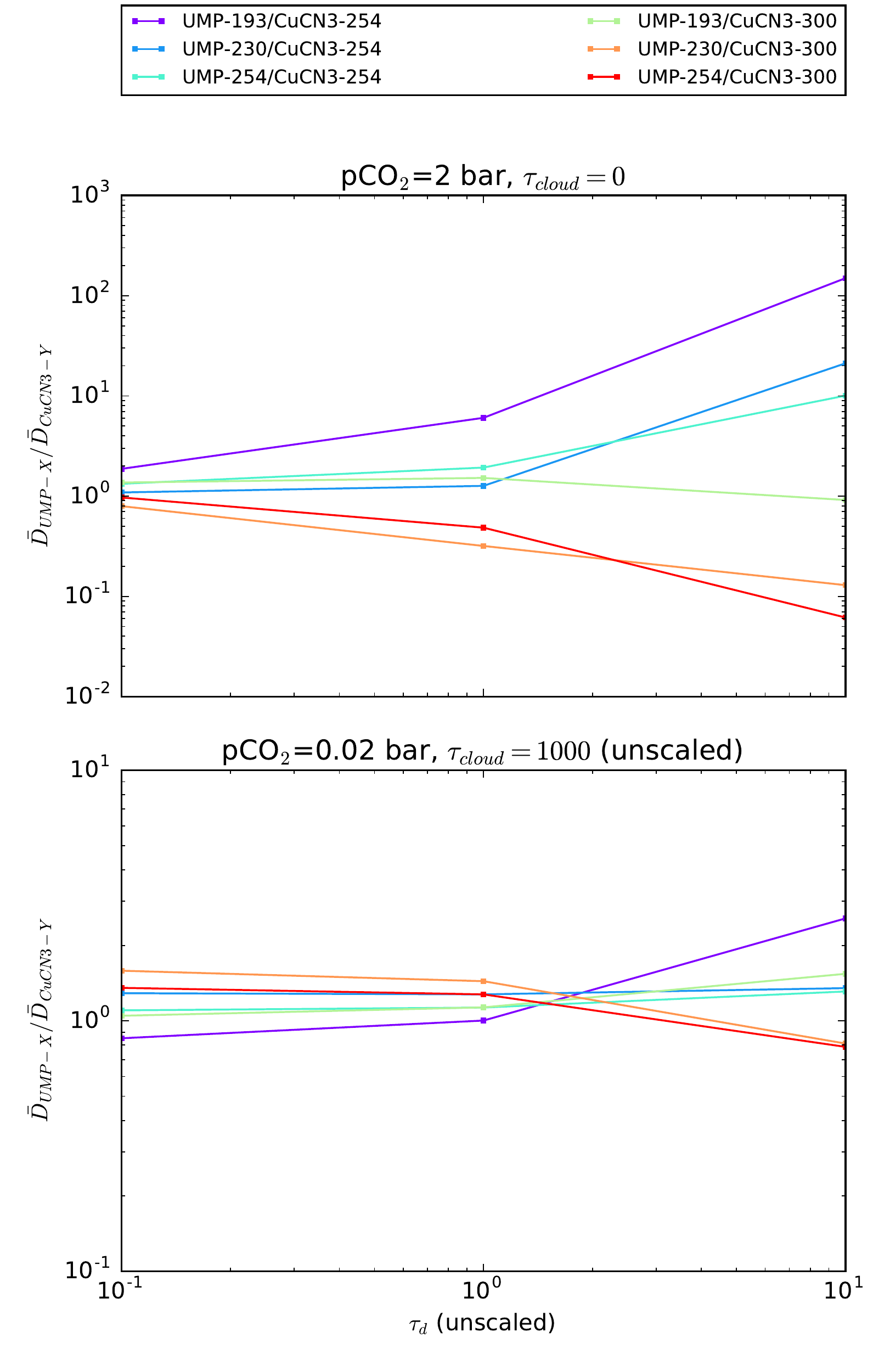}
\caption{Ratio of stressor dose rates UMP-X divided by eustressor dose rates UMP-Y for dusty CO$_2$-H$_2$O atmospheres, as a function of $\tau_d$ ($T_0=250$K, SZA=0, $A$ corresponding to desert). The atmospheres are highly scattering, either because of high pCO$_2$ or thick CO$_2$ clouds.  Lower ratios imply a more favorable environment for abiogenesis as measured by these two photoprocesses. \label{fig:discussion_doses_reldoses_dust}}
\end{figure}

Figure~\ref{fig:discussion_doses_reldoses_dust} presents these calculations. Dust attenuation on its own is relatively flat at CO$_2$-scattering wavelengths, as is cloud attenuation. Consequently dusty/cloudy atmospheres tend to reduce UV fluence in a spectrally flat manner, and favor neither the stressor nor the eustressor pathway. On the other hand, in a thick CO$_2$ atmosphere, the scattering optical depth, and hence amplification of dust absorption, increases as wavelength decreases. Consequently, the dose rate ratio does change with increasing $\tau_d$. However, the direction of the change is sensitive to the value of $\lambda_0$, the ionization threshold for the tricyanocuprate ionization process, and the magnitude is further sensitive to the value of $\lambda_0$ for the UMP cleavage process. We consequently conclude it is possible that thick, dusty atmospheres might be more or less clement for abiogenesis than non-dusty atmospheres, but determining which requires wavelength-dependent measurements of the QYs of these chemical processes in the laboratory. 

\subsection{Highly Reducing Atmospheres}
Recent work suggests that if the reduced gases H$_2$ and/or CH$_4$ were present at elevated levels in a thick ($\sim 1$ bar) atmosphere, collision-induced absorption (CIA) due to the interaction of these gases with CO$_2$ might provide enough greenhouse warming to elevate mean Noachian temperatures above freezing \citep{Ramirez2014, Wordsworth2016b}. \citet{Ramirez2014} found that global mean surface temperatures exceeded 273K for $P_0\geq3$ bar and [H$_2$]$\geq0.05$, with higher concentrations of H$_2$ required for lower $P_0$.  More recently, \citet{Wordsworth2016b} used new ab initio calculations of H$_2$-CO$_2$ and CH$_4$-CO$_2$ CIA to show earlier estimates of the CIA were underestimated, and that 2-10\% levels of CH$_4$ or H$_2$ in a $>1.25$ bar atmosphere could elevate planetary mean temperatures over freezing. While it is unclear if such high reducing conditions can be sustained in the steady state, this scenario remains an intriguing avenue to a Noachian Mars with conditions at least transiently globally clement for liquid water and prebiotic chemistry \citep{Batalha2015, Wordsworth2016b}.

H$_2$ is spectrally inert at UV wavelengths compared to CO$_2$. Based on the constraints on H$_2$ absorption we found \citep{Backx1976, Victor1969}, the contribution of H$_2$ to atmospheric absorption and scattering are negligible for H$_2$ mixing ratios of $0-0.1$. Similarly, CH$_4$ does not absorb at wavelengths longer than 165 nm \citep{Au1993, Chen2004}. Hence, it's absorption is highly degenerate with CO$_2$, and its presence at the levels suggested in \citet{Wordsworth2016b} does not impact the UV surface environment. Photochemically-generated hydrocarbon hazes require CH$_4$/CO$_2$ ratios of $>0.1$, and are consequently expected to be thin or nonexistent in this scenario \citep{DeWitt2009}. Consequently, the UV surface environment in an H$_2$ or CH$_4$-rich atmosphere should be similar to the pCO$_2$=2 bar case discussed in Section~\ref{sec:co2h2o}.

\subsection{Highly-Volcanic Mars  (CO$_2$-H$_2$O-SO$_2$/H$_2$S Atmosphere)}
We have so far considered atmospheres with CO$_2$ and H$_2$O as their dominant photoactive gaseous species. However, other gases have been proposed as significant constituents of the Martian atmosphere. In particular, \citet{Halevy2007} suggest that the lack of massive carbonate deposits on Mars could have been explained if, during epochs of high volcanism on young Mars, SO$_2$ built up to the $\sim1-100$ ppm level. At such levels, \citet{Halevy2007} find that SO$_2$ would supplant CO$_2$ as the agent regulating global chemistry and climate, inhibiting massive carbonate precipitation in the process. \citet{Halevy2014} further argue that enhanced radiative forcing from high SO$_2$ levels could transiently raise mean surface temperatures at the subsolar point (assuming no horizontal heat transport) above the freezing point of water, explaining the observed fluvial features. \citet{Halevy2014} calculate that SO$_2$ mixing ratios $\gtrsim10$ ppm (1 bar atmosphere) could have been possible during, e.g., the emplacement of the Martian volcanic plains. While the impact of SO$_2$ on Martian carbonates and climate remains debated (e.g., \citealt{Niles2013}, \citealt{Kerber2015}), it remains plausible that Noachian Mars may have been characterized by at least transiently high SO$_2$ levels due to higher volcanic outgassing rates.

\begin{figure}[H]
\centering
\includegraphics[width=10 cm, angle=0]{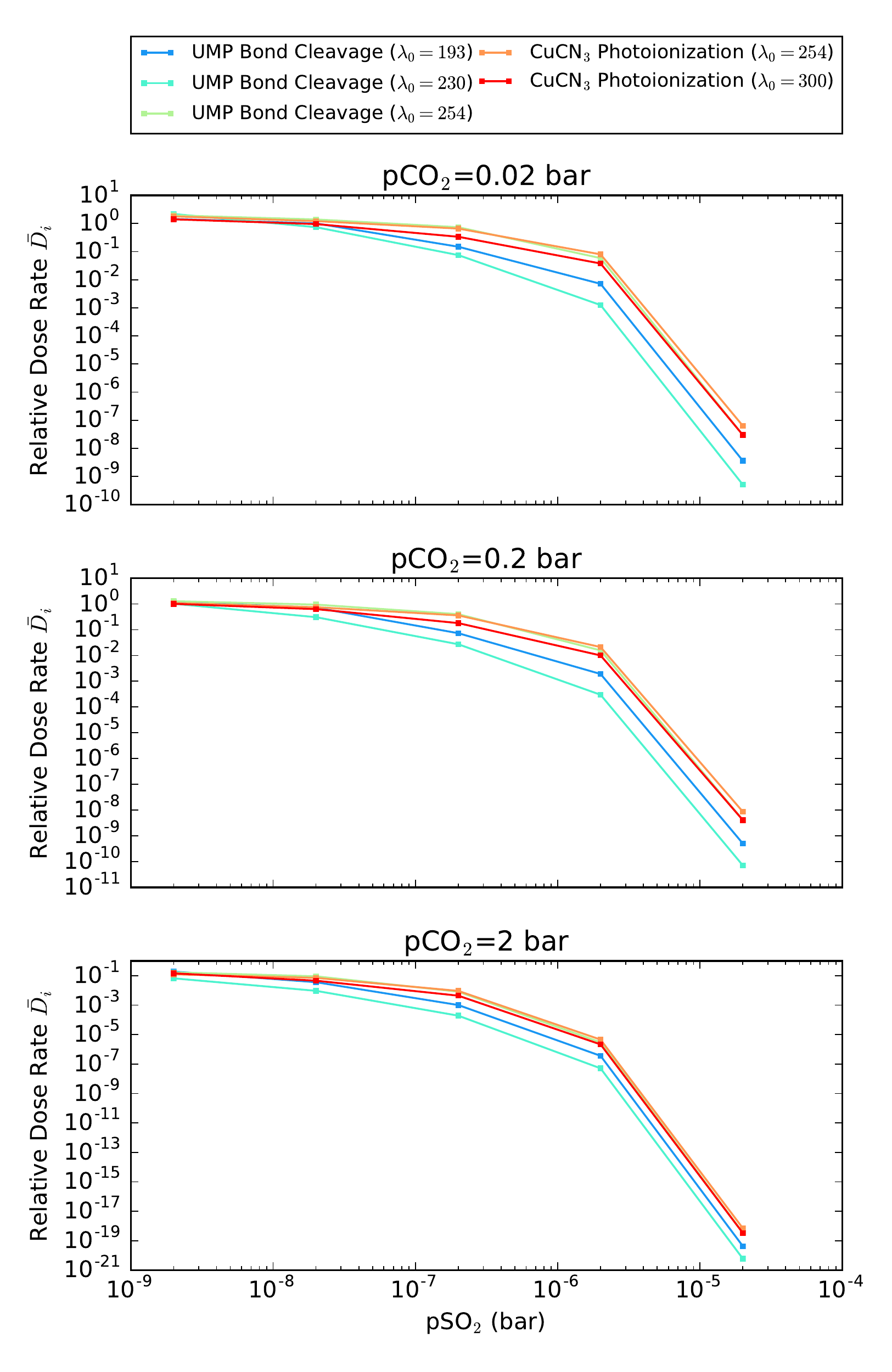}
\caption{UV dose rates $\overline{D}_i$ for CO$_2$-H$_2$O-SO$_2$ atmospheres as a function of pCO$_2$ and pSO$_2$ ($T_0=250$K, SZA=0, $A$ corresponding to desert).\label{fig:discussion_doses_pso2_pco2}}
\end{figure}

\begin{figure}[H]
\centering
\includegraphics[width=10 cm, angle=0]{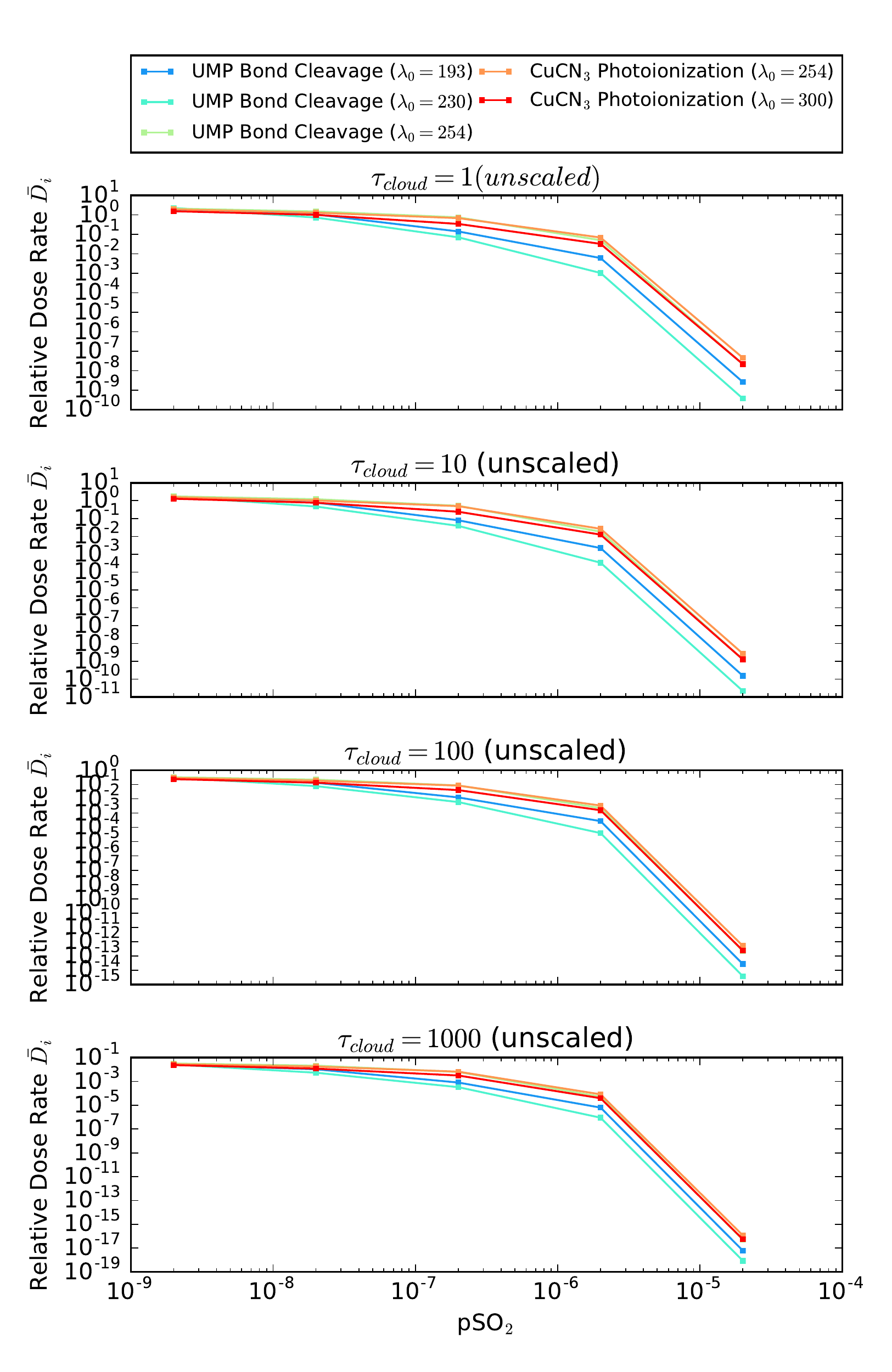}
\caption{UV dose rates $\overline{D}_i$ for CO$_2$-H$_2$O-SO$_2$ atmospheres with a CO$_2$ cloud deck emplaced, as a function of pSO$_2$ and $\tau_{cloud}$ ($T_0=250$K, pCO$_2$=0.02SZA=0, $A$ corresponding to desert).\label{fig:discussion_doses_pso2_clouds}}
\end{figure}

We consequently sought to explore the impact of elevated levels of SO$_2$ on the UV surface environment and hence prebiotic chemistry. Figure~\ref{fig:discussion_doses_pso2_pco2} presents the dose rates calculated for a clear-sky atmosphere with varying pSO$_2$ and pCO$_2$. Figure~\ref{fig:discussion_doses_pso2_clouds} presents the dose rates calculated for an atmosphere with pCO$_2$=0.02 bar (optically thin at scattering wavelengths), but varying levels of CO$_2$ clouds. In both cases, SZA=0 and $A$ corresponds to desert.

SO$_2$ is a far broader, stronger UV absorber than either CO$_2$ or H$_2$O, and consequently its presence can exert a dramatic impact on UV surface radiance and photochemistry rates. As with dust, multiple scattering from other atmospheric constituents can amplify SO$_2$ absorption. For pCO$_2\leq0.2$ bar, pSO$_2\geq2\times10^{-6}$ bar is required to suppress dose rates to $\overline{D}_i<0.1$, whereas for pCO$_2\geq2$ bar, pSO$_2\geq2\times10^{-7}$ bar is sufficient. Similarly, for $\tau_{cloud}\leq10$,  pSO$_2\geq2\times10^{-6}$ bar is required to suppress dose rates to $\overline{D}_i<0.1$, whereas for $\tau_{cloud}\geq100$, pSO$_2\geq2\times10^{-7}$ bar is sufficient. For pSO$_2\geq2\times10^{-5}$ bar, UV-sensitive prebiotic photochemistry is strongly quenched.

\begin{figure}[H]
\centering
\includegraphics[width=10 cm, angle=0]{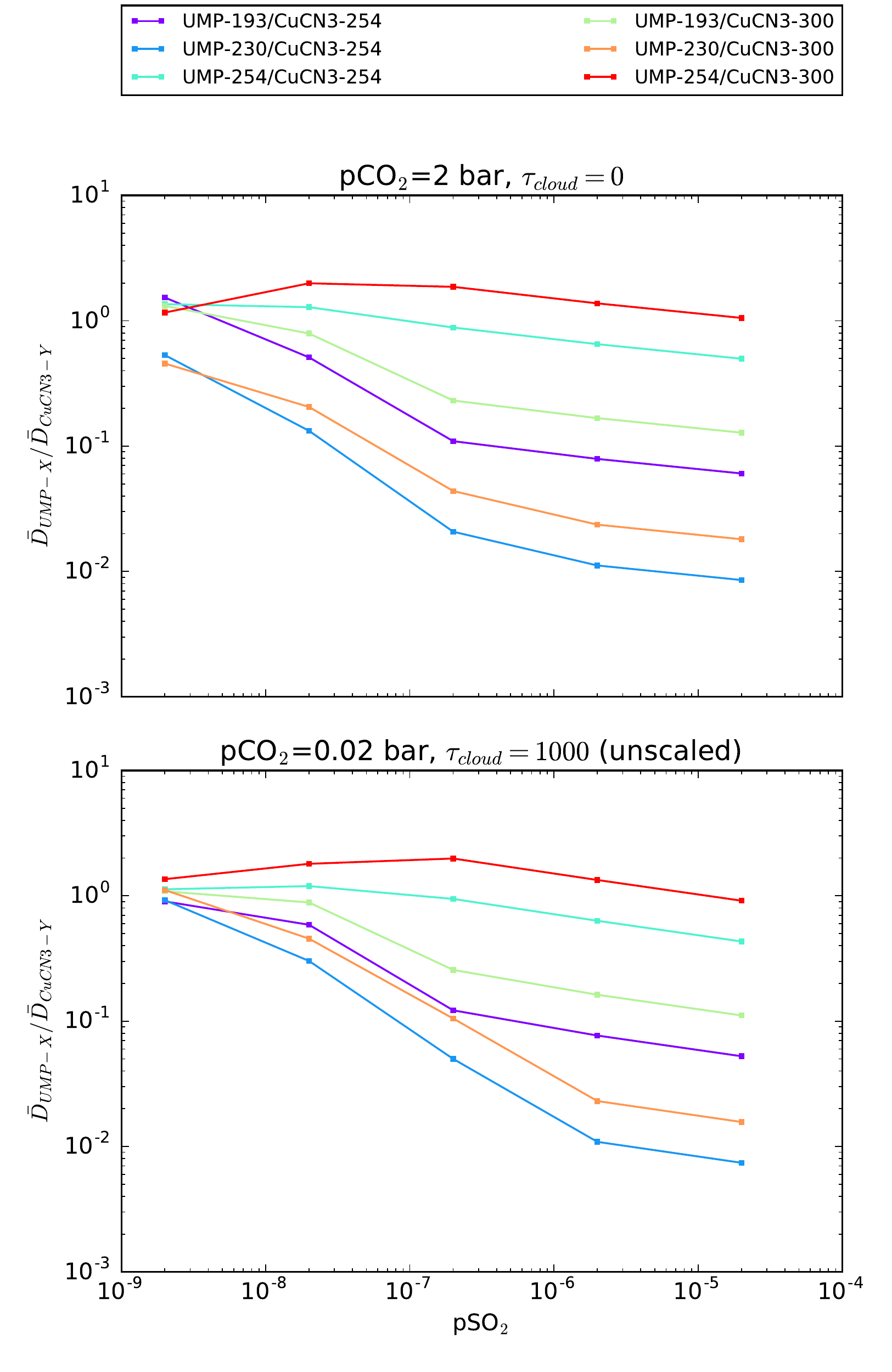}
\caption{Ratio of stressor dose rates UMP-X divided by eustressor dose rates UMP-Y for CO$_2$-H$_2$O-SO$_2$ atmospheres, as a function of pSO$_2$ ($T_0=250$K, SZA=0, $A$ corresponding to desert). The atmospheres are highly scattering, either because of high pCO$_2$ or thick CO$_2$ clouds.  Lower ratios imply a more favorable environment for abiogenesis as measured by these two photoprocesses. \label{fig:discussion_doses_reldoses_pso2}}
\end{figure}

As with dust, we considered the hypothesis that attenuation from SO$_2$ might have a differential impact on the eustressor and stressor pathways. We calculated $\overline{D}_{UMP-X}/\overline{D}_{CuCN3-Y}$ for pCO$_2=2$ bar (no clouds) and $\tau_{cloud}=1000$ (pCO$_2$=0.02 bar) for a broad range of pSO$_2$. This calculation is presented in Figure~\ref{fig:discussion_doses_reldoses_pso2}. We note that regardless of assumption on $\lambda_0$,  as pSO$_2$ increases from $2\times10^{-7}-2\times10^{-5}$ bar, the eustressor pathway is favored  over the stressor pathways, by as much as 2 orders of magnitude (dependent on $\lambda_0$). We conclude that it seems plausible that high-SO$_2$ planetary atmospheres have a UV throughput more clement for abiogensis compared to low-SO$_2$ atmospheres under the assumption that the stressor and eustressor pathways we have identified were important.  However, better measurements of the spectral QY of these photoprocesses is required to confirm and quantify the magnitude of this effect.

While \citet{Halevy2014} focused on the abundance of SO$_2$ in the Martian atmosphere, H$_2$S is emitted in equal proportion by the more reduced Martian mantle \citet{Halevy2007}. \citet{Hu2013} model the atmospheric composition as a function of sulfur emission rate for a 1-bar CO$_2$ atmosphere assuming equipartition of the outgassed sulfur between SO$_2$ and H$_2$S, irradiated by a G2V star at a distance of 1.3 AU. They find H$_2$S concentrations to be even higher than SO$_2$ concentrations, by over an order of magnitude at high S emission rates. H$_2$S is also a stronger, broader UV absorber than CO$_2$ or H$_2$O. Consequently, we sought to explore the impact of elevated levels of H$_2$S on the UV surface environment and prebiotic photochemistry. 

\begin{figure}[H]
\centering
\includegraphics[width=10 cm, angle=0]{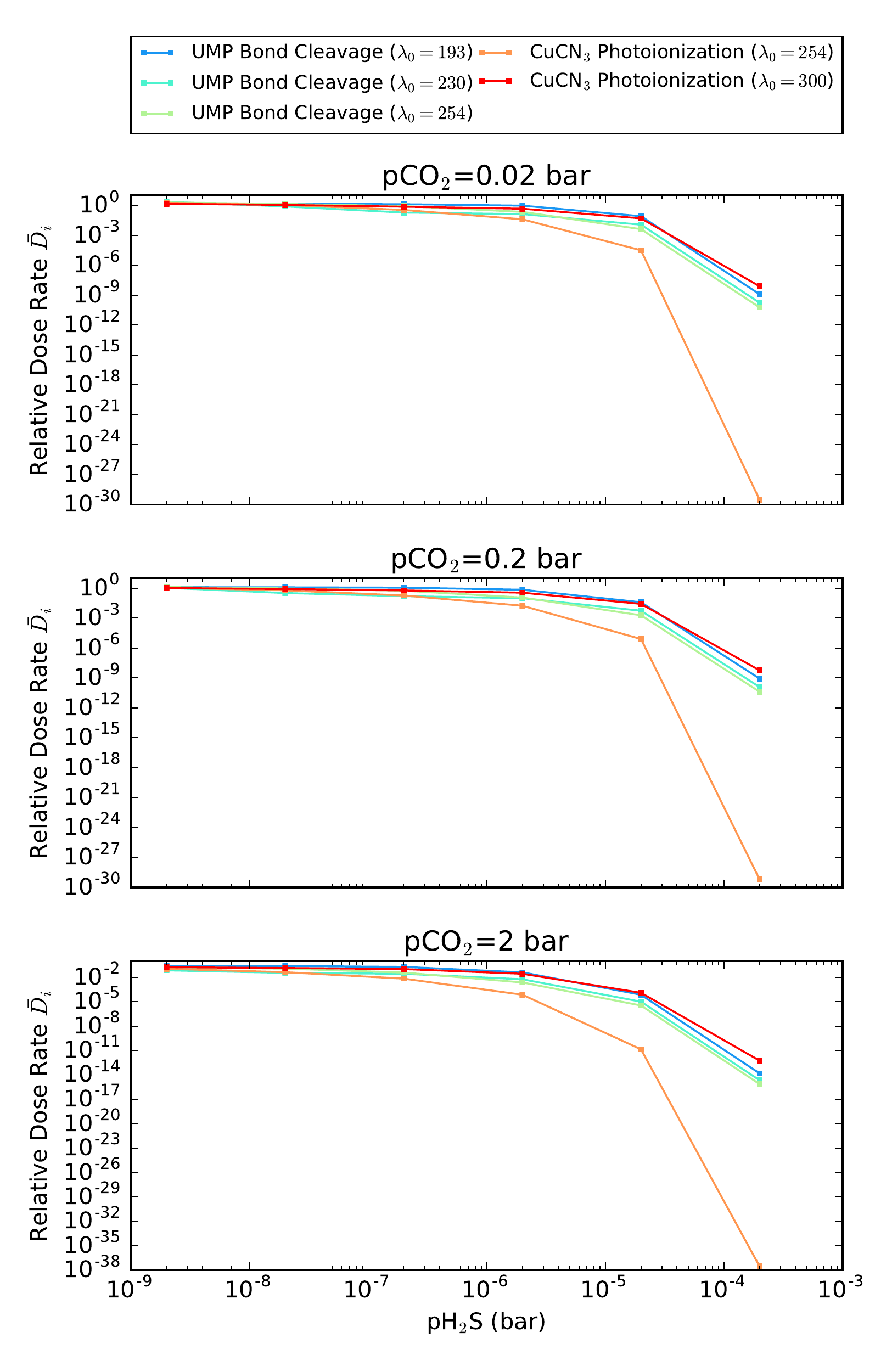}
\caption{UV dose rates $\overline{D}_i$ for CO$_2$-H$_2$O-H$_2$S atmospheres as a function of pCO$_2$ and pH$_2$S ($T_0=250$K, SZA=0, $A$ corresponding to desert).\label{fig:discussion_doses_ph2s_pco2}}
\end{figure}

\begin{figure}[H]
\centering
\includegraphics[width=10 cm, angle=0]{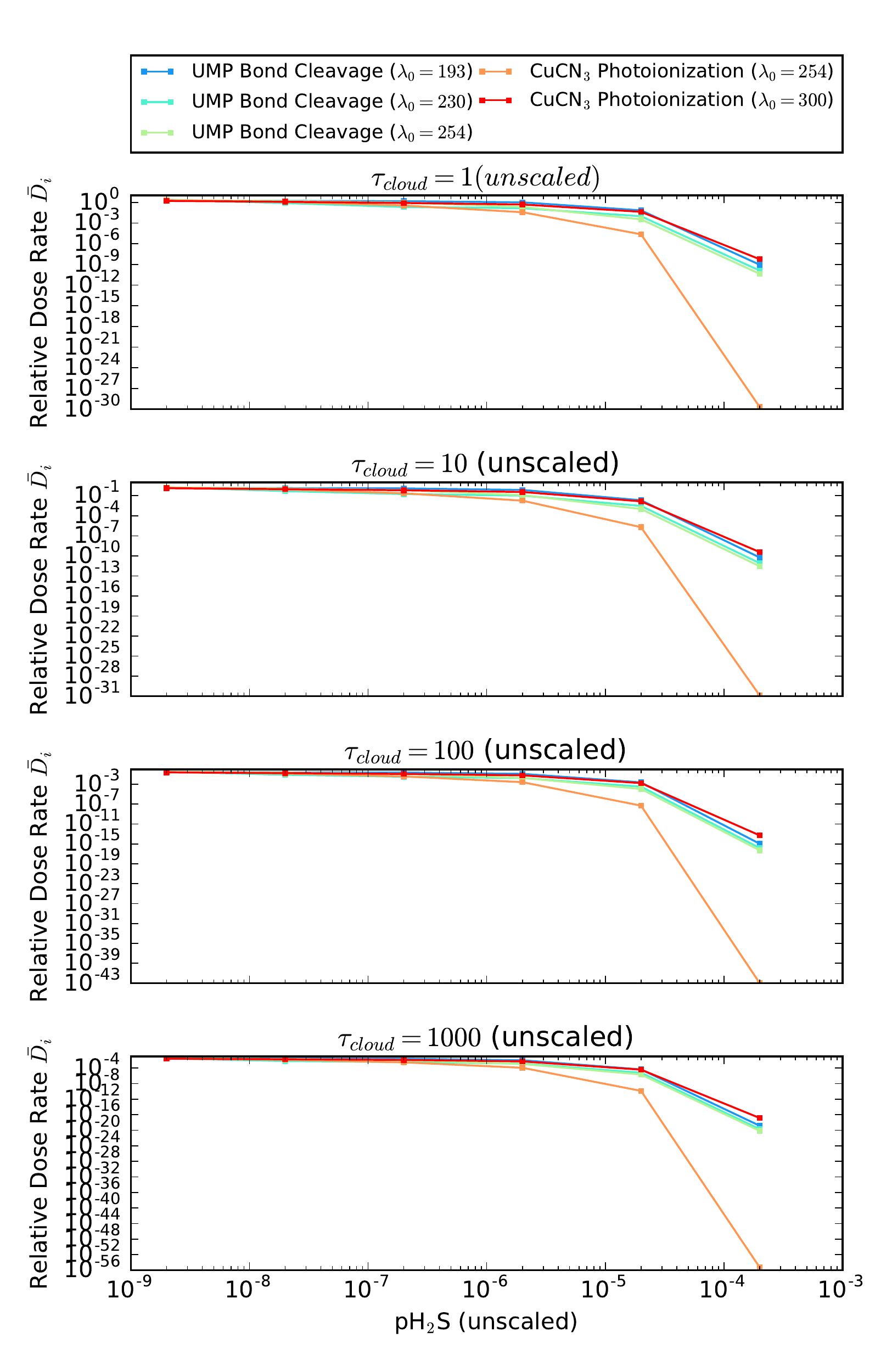}
\caption{UV dose rates $\overline{D}_i$ for CO$_2$-H$_2$O-H$_2$S atmospheres with a CO$_2$ cloud deck emplaced, as a function of pH$_2$S and $\tau_{cloud}$ ($T_0=250$K, pCO$_2$=0.02 bar, SZA=0, $A$ corresponding to desert).\label{fig:discussion_doses_ph2s_clouds}}
\end{figure}

Figure~\ref{fig:discussion_doses_ph2s_pco2} presents the dose rates calculated for a clear-sky atmosphere with varying pH$_2$S and pCO$_2$. Figure~\ref{fig:discussion_doses_ph2s_clouds} presents the dose rates calculated for an atmosphere with pCO$_2$=0.02 bar (optically thin at scattering wavelengths), but varying levels of CO$_2$ clouds. In both cases, SZA=0 and $A$ corresponding to desert. As with SO$_2$ and dust, highly scattering atmospheres can amplify H$_2$S absorption. For pCO$_2\leq0.2$ bar, $\overline{D}_i<0.1$ for pH$_2$S$\geq2\times10^{-5}$ bar, but for pCO$_2\geq2$ bar, $\overline{D}_i<0.1$ for pH$_2$S$\geq2\times10^{-6}$ bar. Similarly, for $\tau_{cloud}\leq10$, $\overline{D}_i<0.1$ for pH$_2$S$\geq2\times10^{-5}$ bar, but for $\tau_{cloud}\geq100$, $\overline{D}_i\lesssim0.1$ for pH$_2$S$\geq2\times10^{-6}$ bar. For pH$_2$S$\geq2\times10^{-4}$ bar, surface photochemistry is strongly quenched regardless of atmospheric state.

\begin{figure}[H]
\centering
\includegraphics[width=10 cm, angle=0]{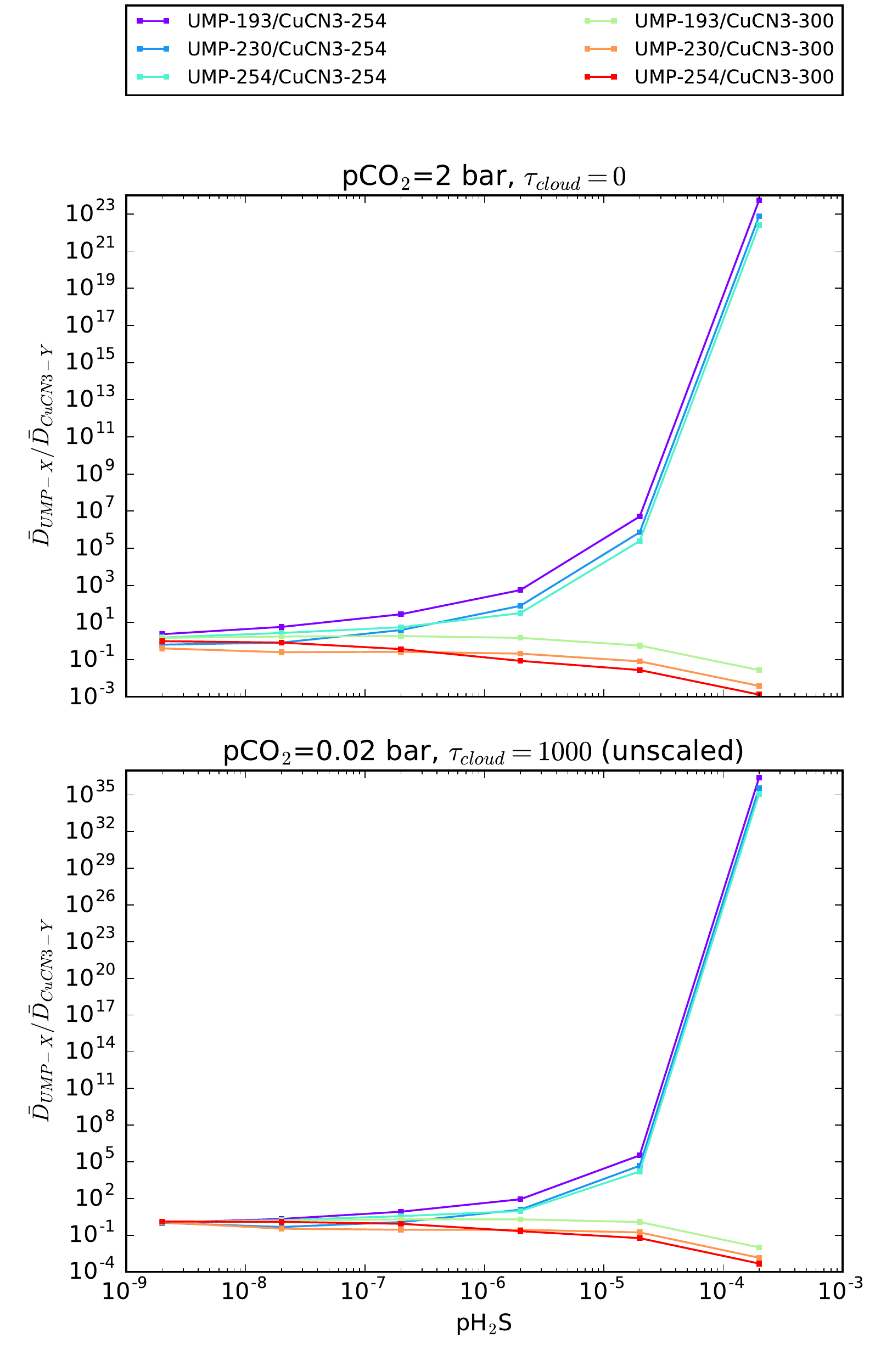}
\caption{Ratio of stressor dose rates UMP-X divided by eustressor dose rates UMP-Y for CO$_2$-H$_2$O-H$_2$S atmospheres, as a function of pH$_2$S ($T_0=250$K, SZA=0, $A$ corresponding to desert). The atmospheres are highly scattering, either because of high pCO$_2$ or thick CO$_2$ clouds.  Lower ratios imply a more favorable environment for abiogenesis as measured by these two photoprocesses. \label{fig:discussion_doses_reldoses_ph2s}}
\end{figure}

We again considered the hypothesis that H$_2$S attenuation might have a differential impact on the stressor and eustressor dose rates. We calculated $\overline{D}_{UMP-X}/\overline{D}_{CuCN3-Y}$ for pCO$_2=2$ bar (no clouds) and $\tau_{cloud}=1000$ (pCO$_2$=0.02 bar) for a broad range of pH$_2$S. This calculation is presented in Figure~\ref{fig:discussion_doses_reldoses_ph2s}. As in the case of dust in a dense CO$_2$ atmosphere, the ratios diverge from 1, but in opposite directions depending on the value assumed for $\lambda_0$ for photoionization of tricyanocuprate. We conclude that attenuation from H$_2$S may well have a differential impact on the stressor and eustressor dose rates, but that assessing which it favors requires better constraints on the QYs of the photoprocesses, especially the photoionization of tricyanocuprate. 

\section{Conclusion\label{sec:conc}}
We have used a two-stream multi-layer radiative transfer model to estimate the UV surface environment on the surface of 3.9 Ga Mars as a function of atmospheric composition, and explored the implications for prebiotic chemistry. Prebiotic photoreaction rates are within an order of magnitude of the terrestrial values for normative clear-sky CO$_2$-H$_2$O atmospheres, in agreement with past work (e.g., \citealt{Cockell2000}) suggesting early Martian and terrestrial atmospheres featured comparable UV environments. In agreement with prior work, we find shortwave radiation to be effectively attenuated by CO$_2$ absorption, with fluence $\lesssim185$ nm removed for pCO$_2\geq 2\times10^{-5}$ bar and fluence $<204$ nm removed for pCO$_2\geq 0.2$ bar. Fluence drops off more slowly in the $>204$ nm regime, where neither CO$_2$ nor H$_2$O absorb. This is a consequence of random walk statistics in highly scattering regimes, and stands in contrast to prior studies\citep{Ronto2003, Cnossen2007} which did not account for multiple scattering. The impact of CO$_2$ and H$_2$O clouds on their own is similarly muted because they too are pure scatterers in the $>204$ nm regime; $\tau_{cloud}\geq100$ is required to significantly affect surficial reaction rates, comparable to but in excess of the highest patchy cloud optical depths predicted by some 3D GCM studies (e.g., \citealt{Wordsworth2013b}). 

While dense atmospheres and cloud decks only modest reduce surface fluence on their own, in concert with other absorbers (dust, SO$_2$, H$_2$S) they can have a dramatic effect on surface fluence and reaction rates, though amplification of the effects of these absorbers. Dust levels of $\tau_d=1$, only a factor of a few higher than that sustained in the modern atmosphere, could suppress prebiotic reaction rates by orders of magnitude for pCO$_2\geq2$bar or $\tau_{cloud}\geq1000$, and dust levels of $\tau_d=10$ would sharply reduce reaction rates independent of the atmospheric state. Similarly, pSO$_2\geq10^{-7}$ bar or pH$_2$S$\geq 2\times10^{-6}$ bars are required to significantly reduce reaction rates for pCO$_2\geq2$bar or $\tau_{cloud}\geq100$, but for less scattering atmospheres, pSO$_2\geq10^{-6}$ bar or pH$_2$S$\geq 2\times10^{-5}$ bar is required. pSO$_2\geq10^{-5}$ bars or pH$_2$S$\geq 2\times10^{-4}$ bars quenches UV-sensitive photochemistry by many orders of magnitude regardless of other atmospheric conditions. 

The absorbers described above have spectrally variable absorption, and prebiotic photochemistry is wavelength-dependent, leading us to speculate whether high abundances of these absorbers, despite suppressing UV-sensitive chemistry generally, might not favor or disfavor eustressor photoprocesses conducive to the origin of life compared to stressor processes that impede life's formation. We compare the relative impact of absorption from dust, SO$_2$, and H$_2$S attenuation on two such stressor (cleavage of the N-glycosidic bond of UMP) and eustressor photoprocesses (production of aquated electrons from CuCN$_{3}^{2-}$. We find that it is possible for high levels of these absorbers to disproportionately favor one or the other of these photoprocesses. In particular, high SO$_2$ levels may create an especially favorable environment for abiogenesis, under the assumption that these photoprocesses were important to abiogenesis (so long as enough fluence reaches the ground to power these reactions). However, the magnitude and direction of this effect is sensitive to assumptions about the QY of these processes. Better characterization of the spectral quantum yields of these processes are required to rule definitively on this question. 

\section{Acknowledgements}
We thank C. Magnani and S. Rugheimer for sharing their data with us, and for many insightful conversations. We thank J. Sutherland and J. Szostak for sharing their insights into prebiotic chemistry. We thank two anonymous referees, whose comments improved this manuscript.

This research has made use of NASA's Astrophysics Data System Bibliographic Services, and the MPI-Mainz UV-VIS Spectral Atlas of Gaseous Molecules.

S. R. and D. D. S. gratefully acknowledge support from the Simons Foundation, grant no. 290360.

\section{Author Disclosure Statement}
The authors declare no competing financial interests.

\clearpage

\bibliography{mars_paper_v4.bib}{}

\clearpage

\appendix
\section{Supplementary Figures\label{sec:appendix_methods}}
\begin{figure}[H]
\centering
\includegraphics[width=10 cm, angle=0]{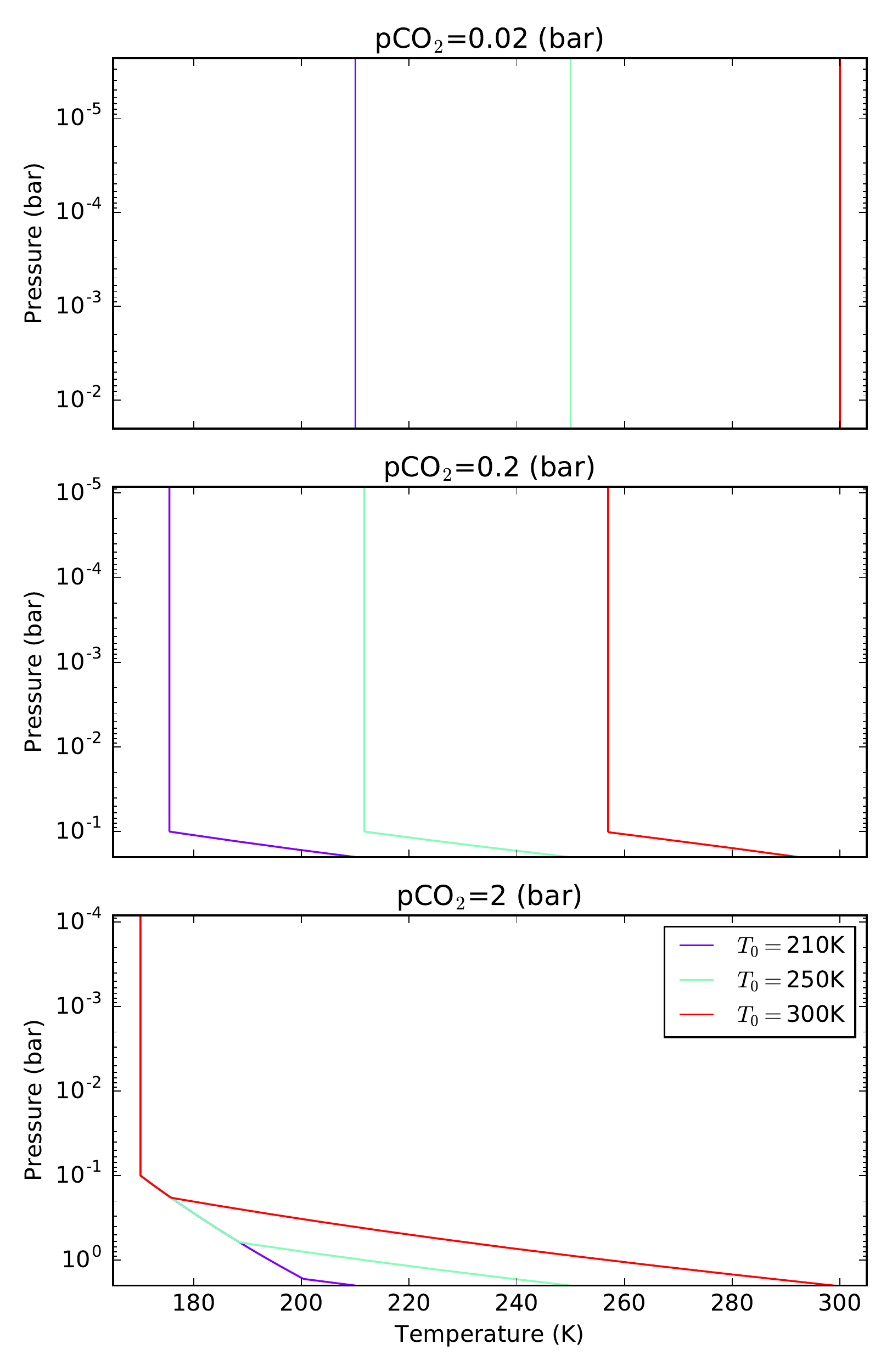}
\caption{Sample T/P profiles for CO$_2$-dominated CO$_2$-H$_2$O atmospheres using the methodology in Section~\ref{sec:atmprofile} for $T_0=210, 250, 300$K and pCO$_2=0.02, 0.2, 2$ bar. The pressure is the total atmospheric pressure (CO$_2$ and H$_2$O). \label{fig:tpprofiles}}
\end{figure}

\begin{figure}[H]
\centering
\includegraphics[width=10 cm, angle=0]{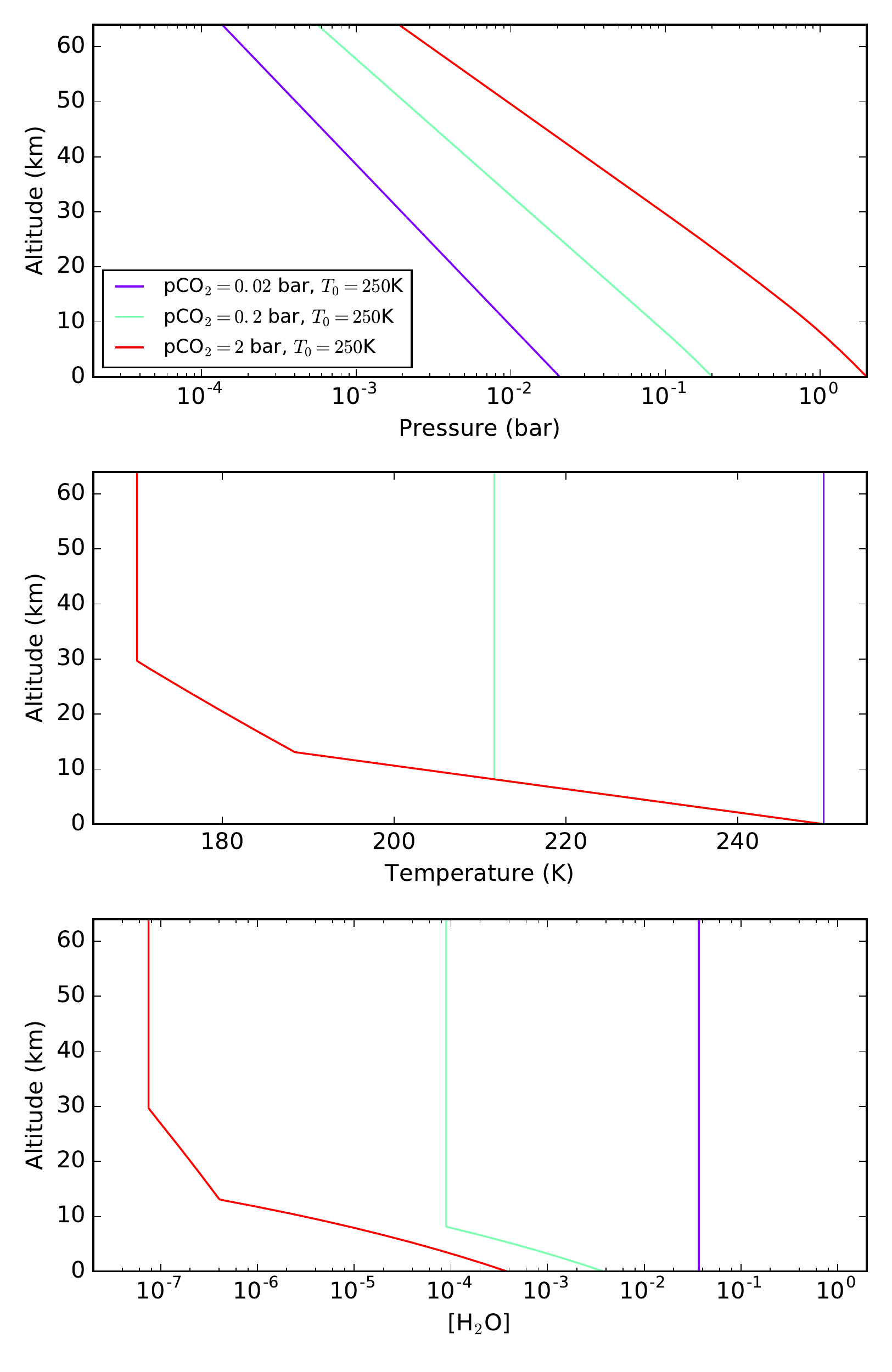}
\caption{Temperature, pressure, and molar concentration of H$_2$O for atmospheres with pCO$_2=0.02, 0.2, 2$ bar and $T_0=250$K. \label{fig:atmoprof_z}}
\end{figure}

This appendix presents sample atmospheric profiles calculated using the methods in Section~\ref{sec:atmprofile}, to illustrate the application of these methods. Figure~\ref{fig:tpprofiles} shows the T/P profiles associated with a sample atmosphere with varying (pCO$_2$, $T_0$), where pCO$_2$ is the surface partial pressure of CO$_2$. Figure~\ref{fig:atmoprof_z} shows the pressure, temperature, and H$_2$O molar concentration altitude provides associated with an atmosphere with $T_0=250$ K and varying pCO$_2$.

\end{document}